%% file: specuPLDI.tex
\documentclass[sigplan,screen]{acmart} 
\usepackage{lipsum}
\usepackage{appendix}

\setcopyright{none}
\acmPrice{}
\acmDOI{}
\acmYear{}
\copyrightyear{}
\acmISBN{}
\acmConference[]{}{}{}
\acmBooktitle{}

\usepackage{booktabs}   %% For formal tables:
\usepackage{subcaption} %% For complex figures with subfigures/subcaptions
\usepackage{amsmath,amssymb,amsfonts}
\usepackage{graphicx}
\usepackage{textcomp}
\usepackage{xcolor}
\usepackage{colortbl}
\usepackage{flushend}
\usepackage{epsfig,color}
\usepackage[utf8]{inputenc}   % input encoding
\usepackage[english]{babel} % US English hyphenation rules
\inputencoding{latin1}
\usepackage{algpseudocode}
\usepackage[ruled]{algorithm}
\usepackage{mdwlist}% Compact list formats \itemize*, \enumerate*
\usepackage{paralist} % for in paragraph enum/item
\usepackage{listings}
\usepackage{multirow}
\usepackage{microtype}
\definecolor{darkblue}{rgb}{0.0, 0.0, 0.55}
\definecolor{darkgreen}{rgb}{0.1, 0.4, 0.1}
\usepackage{hyperref}   % hyperlinks, including DOIs and URLs in bibliography
\usepackage{tikz}
\usepackage{amsthm}

\theoremstyle{definition}

\let\oldnl\nl% Store \nl in \oldnl
\newcommand{\nonl}{\renewcommand{\nl}{\let\nl\oldnl}}% Remove line number for one line

\newcommand{\ignore}[1]{}

\newcommand{\proc}[1]{\textsc{#1}}
\definecolor{mygreen}{rgb}{0,0.6,0}
\definecolor{mygray}{rgb}{0.5,0.5,0.5}
\definecolor{mymauve}{rgb}{0.58,0,0.82}
\definecolor{dkgreen}{rgb}{0,0.6,0}

\definecolor{lightgray}{rgb}{0.85,0.85,0.85}
\definecolor{lightgreen}{rgb}{0.7,0.9,0.7}
\definecolor{lightblue}{rgb}{0.7,0.7,0.9}
\definecolor{lightred}{rgb}{0.9,0.7,0.7}
\algnewcommand\algorithmicforeach{\textbf{for each}}
\algdef{S}[FOR]{ForEach}[1]{\algorithmicforeach\ #1\ \algorithmicdo}

\newcommand\cwnote[1]{\textcolor{red}{{\textbf{Chao Says: #1}}}}

\newcommand\shepherd[1]{#1}

\begin{document}

%% Title information
\title{Abstract Interpretation under Speculative Execution}         %% [Short Title] is optional;
                                        %% when present, will be used in
                                        %% header instead of Full Title.
%\titlenote{with title note}             %% \titlenote is optional;
                                        %% can be repeated if necessary;
                                        %% contents suppressed with 'anonymous'
%\subtitle{Subtitle}                     %% \subtitle is optional
%\subtitlenote{with subtitle note}       %% \subtitlenote is optional;
                                        %% can be repeated if necessary;
                                        %% contents suppressed with 'anonymous'

%% Author information
%% Contents and number of authors suppressed with 'anonymous'.
%% Each author should be introduced by \author, followed by
%% \authornote (optional), \orcid (optional), \affiliation, and
%% \email.
%% An author may have multiple affiliations and/or emails; repeat the
%% appropriate command.
%% Many elements are not rendered, but should be provided for metadata
%% extraction tools.

%% Author with single affiliation.
\author{Meng Wu}
%\authornote{with author1 note}          %% \authornote is optional;
                                        %% can be repeated if necessary
%\orcid{nnnn-nnnn-nnnn-nnnn}             %% \orcid is optional
\affiliation{
%  \position{Position1}
%  \department{Department1}              %% \department is recommended
  \institution{Virginia Tech}            %% \institution is required
%  \streetaddress{Street1 Address1}
  \city{Blacksburg}
  \state{VA}
  \postcode{24060}
  \country{USA}                    %% \country is recommended
}
%\email{mengwu@vt.edu}          %% \email is recommended

%% Author with two affiliations and emails.
\author{Chao Wang}
%\authornote{with author2 note}          %% \authornote is optional;
                                        %% can be repeated if necessary
%\orcid{nnnn-nnnn-nnnn-nnnn}             %% \orcid is optional
\affiliation{
 % \position{Position2a}
%  \department{Computer Science}             %% \department is recommended
  \institution{University of Southern California}           %% \institution is required
%  \streetaddress{941 Bloom Walk}
  \city{Los Angeles}
  \state{CA}
  \postcode{90089}
  \country{USA}                   %% \country is recommended
}
%\email{wang626@usc.edu}         %% \email is recommended

\begin{abstract}
Analyzing the behavior of a program running on a processor that
supports speculative execution is crucial for applications such as
execution time estimation and side channel detection.
Unfortunately, existing static analysis techniques based on abstract
interpretation do not model speculative execution since they focus on
functional properties of a program while speculative execution does
not change the functionality.
%
%However, for applications that care about \emph{non-functional}
%properties, such as cache timing, directly applying these existing
%techniques will produce unsound results.
%
To fill the gap, we propose a method to make abstract interpretation
sound under speculative execution.  There are two contributions.
First, we introduce the notion of \emph{virtual control flow} to
augment instructions that may be speculatively executed and thus
affect subsequent instructions.
Second, to make the analysis efficient, \shepherd{we propose
optimizations to handle merges and loops and to safely bound the
speculative execution depth.}
We have implemented and evaluated the proposed method in a static
cache analysis for execution time estimation and side channel
detection.  Our experiments show that the new method, while guaranteed
to be sound under speculative execution, outperforms state-of-the-art
abstract interpretation techniques that may be unsound.
\end{abstract}




\maketitle

%% Acknowledgments
%\begin{acks}                            %% acks environment is optional
%                                        %% contents suppressed with 'anonymous'
%  %% Commands \grantsponsor{<sponsorID>}{<name>}{<url>} and
%  %% \grantnum[<url>]{<sponsorID>}{<number>} should be used to
%  %% acknowledge financial support and will be used by metadata
%  %% extraction tools.
%  This material is based upon work supported by the
%  \grantsponsor{GS100000001}{National Science
%    Foundation}{http://dx.doi.org/10.13039/100000001} under Grant
%  No.~\grantnum{GS100000001}{nnnnnnn} and Grant
%  No.~\grantnum{GS100000001}{mmmmmmm}.  Any opinions, findings, and
%  conclusions or recommendations expressed in this material are those
%  of the author and do not necessarily reflect the views of the
%  National Science Foundation.
%\end{acks}

\section{Introduction}

Speculative execution~\cite{tomasulo1995efficient} is a feature that
has been implemented by many modern processors.  It allows a processor
to increase the execution speed by exploring certain program paths
ahead of time instead of waiting for the path conditions to be
satisfied.  This is to prevent slower instructions, e.g., memory
accesses, from blocking faster instructions.
For example, when a program reaches a branching instruction,
e.g., \texttt{if(x>5)\{...\}else\{...\}} where the condition depends
on an uncached value of \texttt{x} stored in memory,
a \emph{non-speculative} execution will force the processor to wait,
often for tens or hundreds of clock cycles, until \texttt{x} is loaded
from memory, whereas \emph{speculative} execution allows the processor to
make a prediction of the branching target and then proceed to execute
the predicted branch.
During speculative execution, the processor maintains a checkpoint of
the CPU's register state, which will be used to roll back the changes
if the prediction turns out to be incorrect, i.e., after the value
of \texttt{x} is fetched from memory.  However, if the prediction
turns out to be correct, speculative execution will save time and thus
outperform non-speculative execution.

Speculative execution is designed to be \emph{transparent} to the
program running on the processor; that is, it does not affect the
program semantics, as the rollback ensures that functional properties
are preserved.  This is the reason why, in the past, static analysis
techniques do not model speculative execution.
%For example,
%this is the case even for abstract interpretation based cache analysis
%techniques~\cite{DoychevFKMR13,wilhelm2010static,ferdinand1999cache}.
%
However, recent vulnerabilities such as
Meltdown~\cite{Lipp2018meltdown}, Spectre~\cite{Kocher2018spectre} and
ForeShadow~\cite{vanbulck2018foreshadow} force the community to take
another look because, although speculative
execution preserves the CPU's register state, for performance reasons,
it does not preserve the states of many other components such as the
cache and the pipeline~\cite{GeYCH18,GeYCH19}.
%Thus, upon
%mis-prediction, side effects on these components, which are manifested
%as non-functional properties, cannot be eliminated by rollback.

To see why this may be a problem, consider the cache state that may be
altered by speculative execution and thus affect the timing behavior
of the subsequent non-speculative execution, e.g., cache hits may become
misses, or vice versa.
This is important because an instruction may take only
1-3 clock cycles when there is a cache hit, but tens or even hundreds
of clock cycles when there is a cache miss.
Static analysis is useful in examining the cache related properties of
a program, e.g., to detect information leaks through timing side
channels~\cite{KopfMO12,DoychevFKMR13,BartheKMO14,WuGSW18,SungPW18} or prove that a computation
task always meets the deadline~\cite{ferdinand1999cache,ferdinand1999}.

For side channel detection, in particular, one may want to know if the
program's execution time depends on secret data, e.g., the
cryptographic key, security token, or password.
For deadline estimation, one may want to know the maximum number of
cache misses along program paths, since it corresponds to the
execution time in the worst case.
In both applications, static analysis must be \emph{sound} to be
useful.  By sound, we mean all possible behaviors must be considered.
The reason is because, if the analysis fails to take into
consideration a certain behavior, e.g., a specific execution, it may
miss a bug or security vulnerability, which is not acceptable in
critical applications.

Unfortunately, existing abstract interpretation
techniques \cite{SenS07,DoychevFKMR13,ferdinand1998,ferdinand1999cache}
are unsound under speculative execution.  Instead, these prior works
on abstract interpretation focus more on modeling \emph{non-speculative}
executions, for which numerous techniques have been developed,
including widening/narrowing, chaotic iteration, and efficient
implementations of abstract domains.
Under \emph{speculative} execution, however, none of these techniques
is relevant because the problem is no longer about removing infeasible
paths from the over-approximated analysis, but about preventing real
behaviors from being excluded.  This requires a different set of ideas
from what already exist in the literature.

\begin{figure}%[htb!]
\centering
\includegraphics[width=1.05\linewidth]{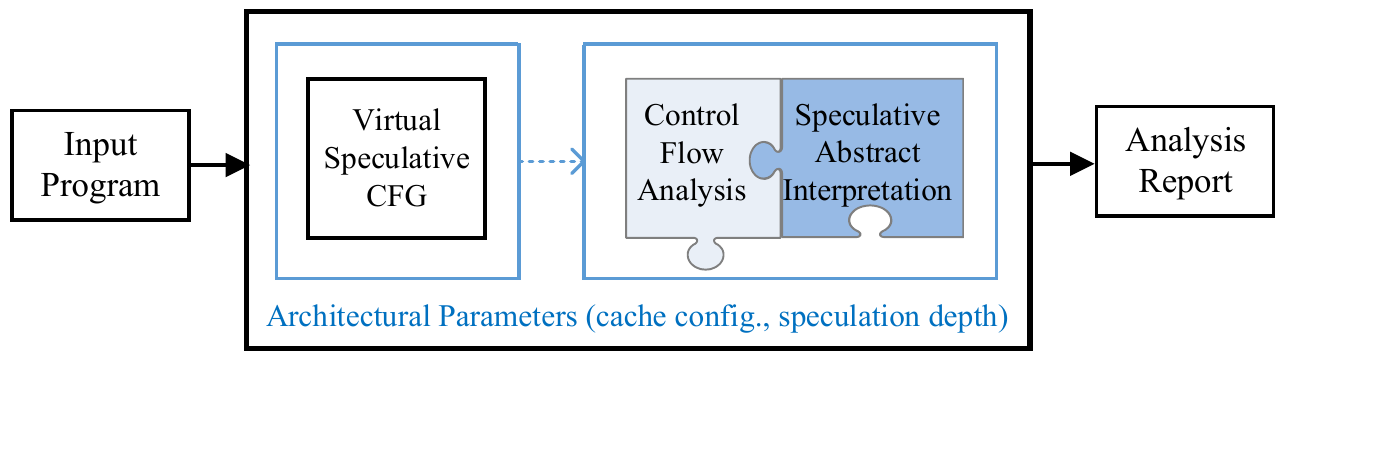}
\vspace{-7ex}
\caption{Our static program analysis framework based on sound abstract interpretation of speculative executions.}
\label{fig:diagram}
%\vspace{-2ex}
\end{figure}

We propose a method for lifting abstract interpretation algorithms so
that they are sound again under speculative execution.
We have developed two techniques.
The first one is a unified way of modeling both non-speculative and
speculative executions using \emph{virtual control flow}.
The second one is redundancy removal, which is crucial for reducing
runtime overhead while maintaining accuracy.

At a high level, the virtual control flow augments the program's CFG
by adding new nodes and edges, e.g., transitions from locations in the
body of one speculatively executed branch to the starting point of the
other branch, to model the rollback upon mis-prediction.
This approach is generally applicable, regardless of how the abstract
state is defined and which algorithm is used to compute the fixed
point.  For example, the abstract state may model side effects on the
cache or pipeline~\cite{schneider1999,schneider1998}, the
non-functional properties to be verified may be timing or power~\cite{EldibWS14tosem,ZhangGSW18,WangSW19,GaoZSW19tosem}, and
the abstract domain may be interval~\cite{cousot1977abstract} or octagonal~\cite{mine2006octagon}.

We have implemented the method in a static cache analysis shown in
Figure~\ref{fig:diagram}, to compute the memory accesses that
correspond to \emph{Must-Hits}. In this context, our method first
computes all possible speculative paths of the program and uses
them to augment the CFG.  Next, it traverses the speculative CFG to
perform abstract interpretation, which computes an abstract (cache)
state for each program location.
To reduce the runtime overhead, it also bounds the depth and number of
speculative executions that abstract interpretation has to consider.
This is possible because, in many cases, the abstract states
are already computed for some location and thus can be used to bound the
speculative executions in other locations. 
Furthermore, we discover that the accuracy of abstract interpretation
is often affected by \emph{when} abstract states from speculative and
non-speculative executions are merged, and we develop a
strategy named ``just-in-time merging'' to minimize the loss of
accuracy.

We have implemented our method in LLVM~\cite{LLVM}, where the
speculative CFG is constructed by an LLVM pass before it is used by
abstract interpretation.
We evaluated it on two types of benchmarks: cryptographic software and
real-time software, where the goal is to detect timing side channels
and to estimate the execution time, respectively.
In both cases, the instruction set architecture is Alpha 21264, with
32-KB fully-associative data cache, 64 bytes per line, and the LRU
replacement policy.
Our experiments show that, compared to existing non-speculative
methods, our method is able to detect significantly
more timing related behaviors, i.e., cache misses and side-channel leaks.
Furthermore, our optimizations are effective in reducing the runtime
overhead while maintaining the accuracy.

To sum up, this paper makes the following contributions:
\begin{itemize}
\item
We show why existing abstract interpretation techniques are unsound for
speculative execution.
\item
We propose a method for lifting existing algorithms to make them sound
for speculative execution.
\item 
We develop optimizations to safely reduce the runtime overhead while
maintaining the accuracy.
\item 
We implement the method and demonstrate its effectiveness on a set of
C programs.
\end{itemize}

The paper is structured as follows.  First, we illustrate the problem
and our solution in Section~\ref{sec:motivation}.  Then, we provide
the technical background in Section~\ref{sec:prelims}, before
presenting our algorithms in
Sections~\ref{sec:baseline}, \ref{sec:approach}
and \ref{sec:optimization}.  We present our experimental results in
Section~\ref{sec:experiment}.  We review the related work in
Section~\ref{sec:related}.  Finally, we give our conclusions in
Section~\ref{sec:conclusion}.

\section{Motivation}
\label{sec:motivation}
	
We illustrate some scenarios in which speculative execution affects
the cache behaviors associated with a program, and explain why such
behaviors are crucial for execution time estimation and side channel
detection.

\subsection{Execution Time Estimation}
	
Figure~\ref{fig:cacheCode2} shows a program that illustrates divergent
cache behaviors under normal and speculative executions as observed
in
practice~\cite{WCET2010Benchmarks,guthaus2001mibench,hpn-ssh,LibTomCrypt,openssh,Tesla}.
Here, we have four variables: \emph{ph}, \emph{l1}, \emph{l2},
and \emph{p}, which are mapped to different cache lines. Suppose the register
value \emph{k} is 0, the load at line 8 will access \emph{ph[0]}.
We assume the cache has 512 lines in total and 64 bytes per 
line.  We also assume the cache
is fully associative, meaning any variable may be mapped to a different 
line.
The place holder variable \textit{ph} is mapped to the first 510 lines (line 3); in practice, \textit{ph} may correspond to an
assorted set of program variables.  Each of the remaining
variables, \textit{l1}, \textit{l2} and \textit{p}, may be mapped to a
cache line.

\begin{figure}
\centering
\begin{minipage}{.45\textwidth}
\begin{lstlisting}[ numbers=left, emph={char, for, load, if, else, int, reg}]
	char ph[64*510], l1[64], l2[64], p;
	reg char k; 
	for(reg int i=0;i<64*510; i+=64) load ph[i]; 
	if(p==0)
		load l1[0];
	else
		load l2[0];
	load ph[k];	
\end{lstlisting}
\end{minipage}
\caption{Example program for timing side channel.}
\label{fig:cacheCode2}
\end{figure}

%\begin{figure}
%\centering
%\begin{minipage}{.45\textwidth}
%\begin{lstlisting}[numbers=left, emph={char, for, load, if, else, reg, int}]
%char ph[64*510], l1[64], l2[64], p; 
%for(reg int i=0;i<64*510; i+=64) load ph[i]; 
%if(p==0)
%  load l1[0];
%else
%  load l2[0];
%load ph[0];
%\end{lstlisting}
%\end{minipage}
%\caption{Example program for execution time estimation.}
%\label{fig:cacheCode1}
%\end{figure}
%	

\begin{figure*}
\centering
\includegraphics[scale=.65]{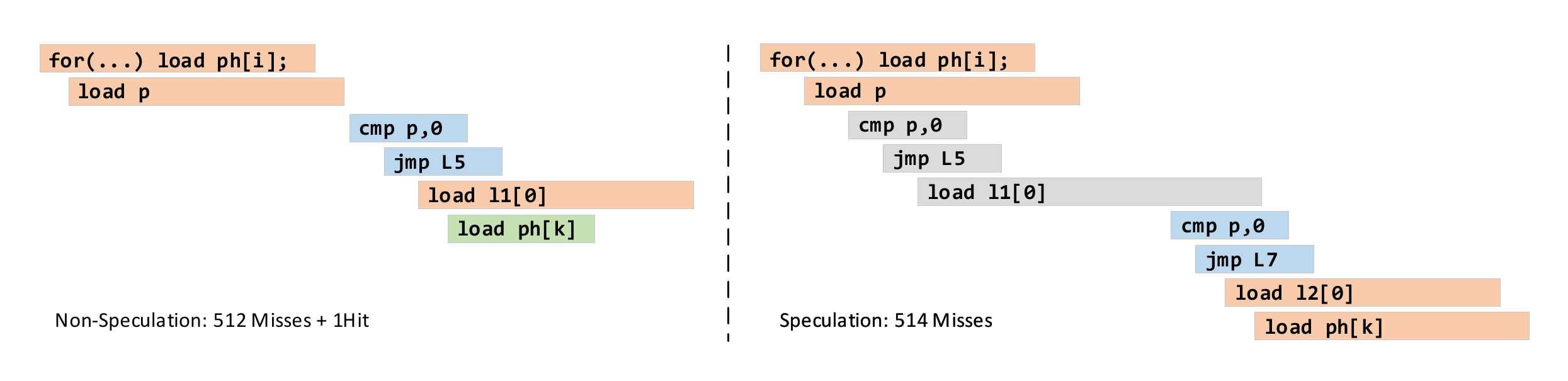}

\vspace{-2ex}
\caption{Pipelined execution trace for program in Figure~\ref{fig:cacheCode2}} 
\label{fig:pipeline}
\end{figure*}

Depending on the branching condition, either \emph{l1} or \emph{l2}
may be loaded to the cache, but both will
result in 512 cache misses.  As shown on the left-hand side of
Figure~\ref{fig:pipeline}, the statement at line 8,
accessing \emph{ph[0]}, is always a hit because the content is already
in the cache.

However, under speculative execution, upon reaching the \emph{if-else}
statement, the CPU needs to load \emph{p} from memory.  Due to a cache
miss, it performs a speculative execution of the branch \texttt{(p==0)}.
If the branch prediction is incorrect and the CPU has to roll back the
speculative execution, there will be 514 cache misses (among which 513
cache misses are observable from outside of the CPU) as shown by the
right-hand-side trace in Figure~\ref{fig:pipeline}.

In this case, the program first speculatively executes the
\emph{then}-branch and loads \texttt{l1} into the cache, and  then rolls back to 
take the \emph{else}-branch and loads \texttt{l2}. Although the
functional side-effects of executing the \emph{then}-branch are
eliminated by the rollback mechanism, \texttt{l1} is already in the
cache.  Since the cache has only 512 lines, following the LRU
replacement policy, the first line associated with \textit{ph[0]} is
evicted.  This is why the subsequent access to \emph{ph[0]} will be a
cache miss.

For execution time estimation, the non-speculative execution will
lead to 512 cache misses plus 1 cache hit, whereas the speculative
execution will lead to 513 observable cache misses (and a speculative
cache miss masked by the pipeline). The additional cache miss is
important because it will cause a significant delay in the execution
time.
The message from this example is as follows: if a static analysis is
not sound in modeling speculative execution, it may underestimate the
worst-case execution time and produce a bogus proof that the
computation task meets its deadline.

\subsection{Side Channel Detection}

We use Figure~\ref{fig:cacheCode2} again to illustrate a timing side
channel made possible by speculative execution. That is, the attacker,
by measuring the execution time of a program, may deduce information
of the secret data.
This time, we assume the variable $k$ stores the secret data, e.g., a
cryptographic key, and the value of $k$ is used as an index to access
an \emph{S-Box}-like array named  \textit{ph}. If the time taken by the
access varies with respect to $k$, there is an information leak.

In a non-speculative execution, there cannot be leaks in
Figure~\ref{fig:cacheCode2} because, for all paths and values
of \emph{k}, the number of cache misses remains the same.
In particular, accessing \textit{ph[k]} is leak-free because the array
is loaded to cache at line 3, and executing either branch at lines 5
and 7 will not evict it.  However, similar to what we have observed in
the execution time estimation example, speculative execution may
execute one of the two branches first, and then roll back to execute
the other branch.  Since the memory locations associated with both
branches must be accessed, which add up to more than 512 cache lines,
some of the cache lines associated with \textit{ph} will be evicted.
Therefore, the subsequent \textit{load} (at line~8) may be a cache
miss.  The difference in execution time may be observed by the
attacker and used to deduce information of the secret $k$: whether the
last statement leads to a cache miss depends on the value of \emph{k}.

\subsection{Technical Challenges}

The above two examples illustrate the need to soundly model
speculative execution.
However, there are several challenges. 
The first one is to model the cache state of a program during
speculative execution without drastically altering the 
abstract interpretation algorithm.
The second challenge is to judiciously merge abstract states computed
from normal and speculative executions, since \emph{when}
and \emph{how} to merge them drastically affect the accuracy of the
fixed-point computation.
Furthermore, since a speculative execution may be rolled back at any
moment, the number of scenarios is exponential in the number of
speculatively executed instructions.  If we have to enumerate, the
analysis time will be prohibitively long.  Therefore, we group scenarios 
into equivalence classes, based on which we perform reduction to
balance the performance and accuracy.

In the remainder of this paper, we will present our solutions in
detail.

\section{Preliminaries}
\label{sec:prelims}

We review the basics of abstract interpretation, as well as the cache,
branch prediction, and speculative execution.

\subsection{Abstract Interpretation}

Abstract interpretation~\cite{cousot1977abstract} is a static analysis
framework that considers all paths and inputs to obtain a sound
over-approximation of the state at every program location~\cite{KusanoW16,KusanoW17,SungKW17}.
For efficiency reasons, the state is kept \emph{abstract} and often
represented by a set of constraints in a certain \emph{abstract
domain}.  
For example, in the interval domain, each constraint is of
the form $\mathit{lb} \leq x \leq \mathit{ub}$, where $x$ is a
variable and $\mathit{lb}$,$\mathit{ub}$ are the lower and upper
bounds.  The join of two states, $s_1 = \mathit{lb_1} \leq
x \leq \mathit{ub_1}$ and $s_2 = \mathit{lb_2} \leq
x \leq \mathit{ub_2}$, is defined as $s_1\sqcup s_2
= \mathit{min(lb_1,lb_2)} \leq x \leq \mathit{max(ub_1,ub_2)}$.  Here,
$\sqcup$ denotes the \emph{join} operator, which returns an
over-approximation of the set union.
If, for example, the polyhedral abstract domain is used, a constraint
will be a linear equation and the \emph{join} operator may be the
convex hull.

The purpose of restricting the representation of states to an abstract
domain is to reduce the computational overhead.  Although various
abstract domains may be plugged in, the underlying fixed-point
computation remains the same.  The fixed-point of states are computed
on the program's control flow graph (CFG). Without loss of generality,
we assume the CFG has a unique entry node and a unique exit node.
Inside the CFG, nodes are associated with instructions or basic
blocks of instructions, whereas edges represent the control flows,
guarded by conditional expressions.

Let $\proc{Transfer}: S \times \mathit{INST} \rightarrow S$ be the
transfer function, which takes a state $s\in S$ and an instruction
$\mathit{inst}\in \mathit{INST}$ as input, and returns the new state
$s' = \proc{Transfer}(s,\mathit{inst})$ as output. $s'$ is the result
of executing $inst$ in state $s$.

\begin{algorithm}[t!]
\caption{Abstract interpretation based static analysis.}
\label{alg:baseline}
\footnotesize
\begin{algorithmic}[1]
%	\Function {StaticAnalysis}{$CFG$}
	\State Initialize $S[n]$ to $\top$ if $n =$ \Call{Entry}{$CFG$}, and to $\bot$ otherwise 
	\State $WL \gets$ \Call{Entry}{$CFG$}
	\While{$\exists$ $n \in WL$ }
	        \State $WL \gets WL \setminus \{n\}$
	        \State $s' \gets $ \Call{Transfer}{$S[n]$, $inst_n$}
		\ForAll{$n' \in $ \Call{Successors}{$CFG$, $n$}}
		\If{$s' \not \sqsubseteq S[n'])$}
			\State $ S[n'] \gets$ $s[n'] \sqcup s'$
			\State $WL \gets WL \cup \{n'\}$
		\EndIf
		\EndFor
	\EndWhile
%	\EndFunction
\end{algorithmic}
\end{algorithm}

Algorithm~\ref{alg:baseline} shows a generic procedure that returns,
for each CFG node $n$, an abstract state $S[n]$ as output.
$S[n]$ is supposed to be a sound over-approximation of all the
possible states at $n$, regardless of the input values or paths taken
to reach $n$.
Initially, $S[n]$ is $\top$ (tautology) for the entry node but $\bot$
(empty) for all other CFG nodes.
The remaining part of the procedure is a standard worklist-based
algorithm for computing the fixed point~\cite{nielson1999}: starting
from the entry node, it computes the states of the successor nodes
($n'$) based on the transfer function.
To ensure convergence, e.g., when the program has loops or is
otherwise non-terminating, a \emph{widening} operator ($\nabla$) is
needed in addition to \emph{join} ($\sqcup$).  However, for brevity,
we omit the details; for a complete introduction, refer
to~\cite{cousot1977abstract,mine2006octagon}.

The actual definitions of abstract state $S$ and transfer function
$\proc{Transfer}$ depend on the application.  In this work, we are
concerned with the cache state corresponding to a program.  We will
present our definitions in Section~\ref{sec:baseline}.

\subsection{Cache and Speculative Execution}

Cache is a type of small but fast storage to hold frequently used data
so that they do not need to be fetched from or stored to the large but
slow memory every time.
Although this work focuses on the data cache, which is more relevant
to our applications, the underlying technique can be extended to the
instruction cache as well.

In a typical CPU, e.g., an Intel processor~\cite{Intel}, instructions
are fetched from memory and decoded continuously before they are sent
to the scheduler for execution.  Executing an instruction involves
multiple units; speculative execution~\cite{tomasulo1995efficient} is
an optimization that efficiently utilizes these execution units.
During speculative execution, instructions are scheduled in a pipeline
as soon as the required execution units are available; for example,
while an instruction is waiting for data to be fetched from memory,
subsequent instructions may be executed, as long as the program
semantics remains the same to observers from outside of the CPU.

Things become complicated when there are branches, however, since the
branch prediction unit must make a guess on which branch target to
execute. Instructions in the predicted branch will be executed while
the branch condition is being evaluated, and will be committed only
after the prediction is confirmed to be correct.  Upon misprediction,
however, the result of speculative execution will be discarded and the
execution will be redirected to the correct branch.

The reorder buffer inside the execution unit, among others, is
responsible for this \emph{rollback}: upon a branch mis-prediction, it
will not perform register retiring as in a normal execution; instead,
it will flush out the affected registers, before restoring the CPU to
a previously saved state.

\ignore{
\begin{figure*}
\centering
\includegraphics[scale=.8]{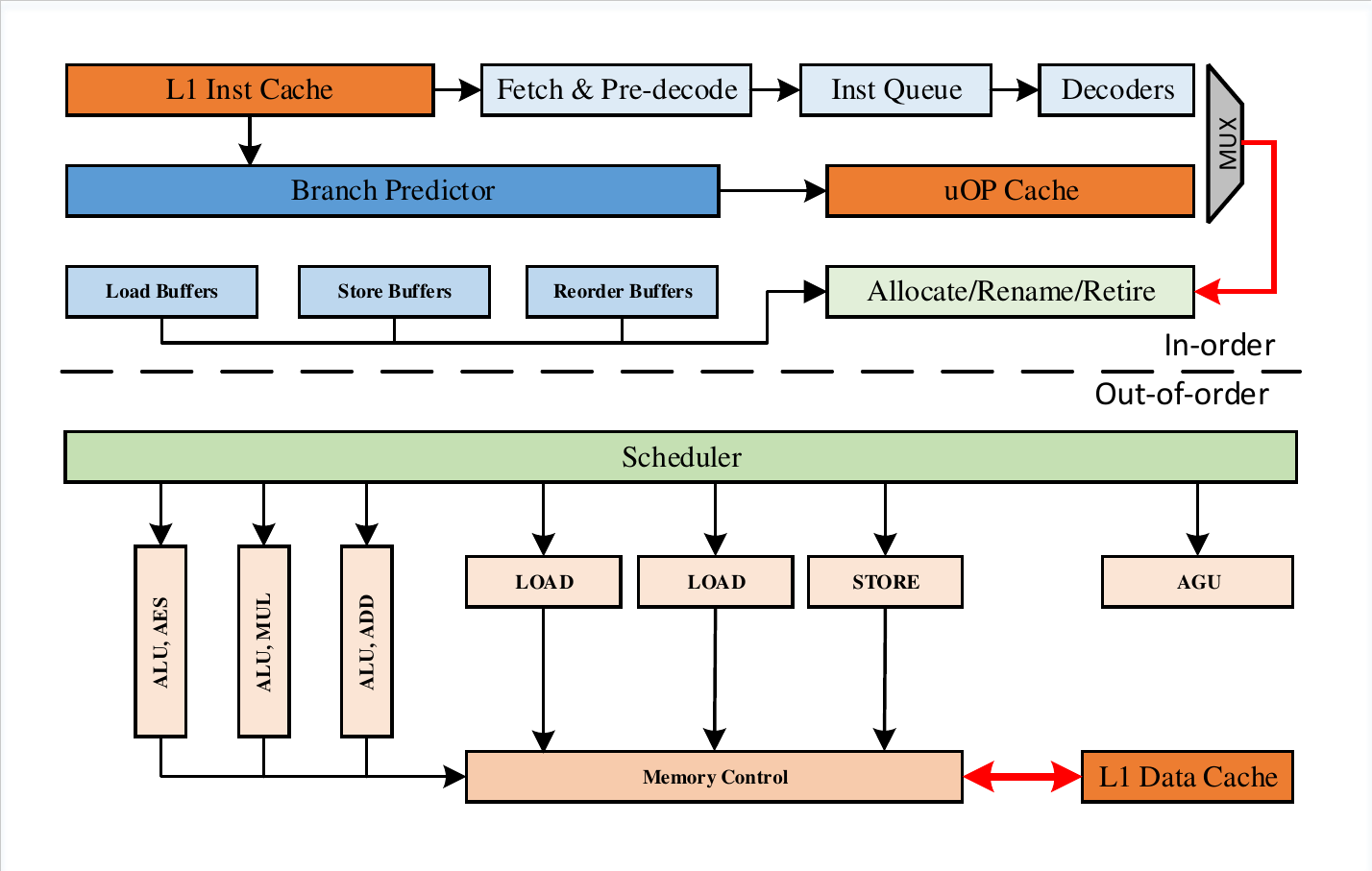}
\caption{Simplified illustration of a single core CPU's micro-architecture.} %Note: copy from meltdown paper!!
\cwnote{Should we remove the two blocks related to instruction cache?}
\label{fig:cpu_arch}
\end{figure*}
}

The branch predictor also plays an important role in speculative
execution since its accuracy is directly related to the performance of
the CPU.
%The popular strategies used in branch prediction fall into
%three categories: static, dynamic, and hybrid.
%%
%Static branch prediction~\cite{hennessy2011} is performed at compile
%time, often based on static program analysis or data collected through
%profiling.  As such, they are less accurate than dynamic branch
%prediction~\cite{cheng2000,yeh1991two,yeh1992alternative}, which has
%access to runtime information such as the branch history and branch
%correlations.
%%
%Hybrid approaches~\cite{mcfarling1993combining} aim to take advantage
%of both static and dynamic information, and sometimes use techniques
%such as neural branch
%prediction~\cite{vintan1999towards,teran2016perceptron} to improve the
%performance.
%
However, regardless of the underlying strategies~\cite{jimenez2001dynamic,yeh1991two,vintan1999towards}, when a branch
prediction turns out to be incorrect, the speculatively executed
instructions may leave side-effects on the states of other system
components, including the cache.
In this work, we are concerned with modeling of such side-effects in
abstract interpretation.

\section{Static Cache Analysis}
\label{sec:baseline}

In this section, we present our instantiation of the baseline abstract
interpretation algorithm.  The goal is to decide, at each program
location, whether a memory access always results in a cache hit.  
Previously, such \emph{must-hit} analyses were used in execution time
estimation~\cite{ferdinand1999,ferdinand1998} and side channel
mitigation~\cite{DoychevFKMR13,BartheKMO14,WuGSW18}; however, they did
not model speculative execution.

\subsection{The Abstract State}

Let $V = \{v_1,...,v_n\}$ be the set of program variables stored in
memory.  Each variable $v\in V$ may be mapped to a cache line.  Let
the cache be fully associative with the LRU replacement policy, which
means a variable $v\in V$ may be mapped to any cache line and, if
there is not enough space, the least recently used (LRU) variable will
be evicted from the cache.  Assume that $N$ is the total number of
cache lines, we can define the age of each variable
$v\in V$, denoted $\mathit{Age}(v)$, which is an integer ranging from $1$ to $N+1$.
Here, $\mathit{Age}(v)=1$ means $v$ resides in the most recently used
line, $\mathit{Age}(v)=N$ means $v$ resides in the least recently used
cache line, and $\mathit{Age}(v) = N+1$ means $v$ is outside of the
cache.

The cache state $S$ associated with the entire program is defined as
$S = \langle Age(v_1), \ldots, Age(v_n) \rangle$.
In this context, a \emph{Must-Hit} analysis needs to compute, at each
program location, an upper bound of $Age(v)$.
If the upper bound is less than or equal to $N$, then $v$ must be in
the cache.  Otherwise, it is possible that $v$ may be outside of the
cache.

\subsection{The Transfer Function}
\label{sec:transfer}

Let $\proc{Transfer}(S,inst)$ be the transfer function that models the
impact of executing $inst$ in the cache state $S$: given the current
state $S
= \langle \mathit{Age}(v_1),\ldots,\mathit{Age}(v_n) \rangle$, it
returns a new state $S'
= \langle \mathit{Age'}(v_1),\ldots,\mathit{Age'}(v_n) \rangle$.
If $inst$ does not access memory at all, then $S' = S$.
Otherwise, assume that $v\in V$ is the variable being accessed in
$inst$, and we compute the new state $S'$ as follows:
\begin{itemize}
\item 
For the accessed variable $v$, set $\mathit{Age}'(v) = 1$ in $S'$.
\item 
\shepherd{For variable $u \in V$ whose age may be younger than $v$ in $S$,
increment the age of $u$; that is, \\ $\mathit{Age}(u) < \mathit{Age}(v) \rightarrow \mathit{Age'}(u) = \mathit{Age}(u) +1$.}
\item 
For any other variable $w \in V$, set $\mathit{Age'}(w)
= \mathit{Age}(w)$.
\end{itemize}
Given the function $\proc{Transfer}$ for an instruction, we
define it for a sequence of instructions $Insts = \{inst_0,
inst_1, ...inst_n\}$ as follows: $  \proc{Transfer}(S,Insts) = $
\[
  \proc{Transfer}( \ldots (\proc{Transfer}(S,inst_0),inst_1),\ldots, inst_n).
\]

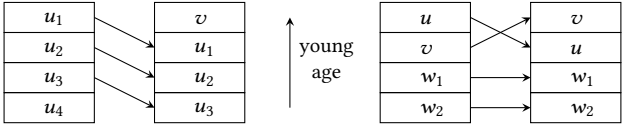
\begin{figure}%[htb!]
\centering
\scalebox{1}{\input{fig/cacheupdate}}
\caption{Transfer of the cache state under the LRU policy.}
\label{fig:cacheupdate}
% \cwnote{Is this (fully associative cache) the most general case (subsuming all other types of cache)?}
\end{figure}

Figure~\ref{fig:cacheupdate} shows two examples.  The left-hand-side
example illustrates the access of $v$, which is not yet loaded into
the cache. After the access, $Age(v)=1$, meaning $v$ is loaded to the
youngest cache line.  Furthermore, the ages of all other lines
increase by 1.  Since $Age(u_4)>4$, the variable $u_4$ is evicted from
the cache.

In the right-hand-side example, however, $v$ is in the cache prior to
the execution of the instruction.  Thus, existing cache lines fall
into two categories.  For the variable ($u$) whose age used to be younger
than that of $v$, the age increases by 1. For the variables ($w_1$
and $w_2$) whose ages used to be older than that of $v$, the ages
remain the same.

\subsection{The Join Operator}
\label{sec:join}
	
\begin{figure}%[htb!]
\centering
\scalebox{1}{\input{fig/cachejoin}}
\caption{Join of two states at a control-flow merge point.}
\label{fig:cachejoin}
\end{figure}
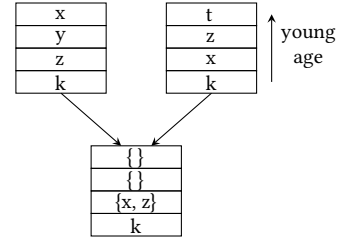  

For efficiency reasons, states computed along two program paths
are joined together at the control-flow merge point, to avoid
creating an exponential number of states.  In the baseline abstract
interpretation algorithm, the join operator ($\sqcup$) always
maintains a single cache state in the result, regardless of how many
states are joined.

Figure~\ref{fig:cachejoin} illustrates a join operator that 
computes, for each variable $v\in V$, the maximum possible age.  For
example, the ages of variable $x$ were 1 in the state on the left and
3 in the state on the right; thus, the age of $x$ after join is 3.
Similarly, for variable $z$, the ages were 3 and 1; thus, the age
after join is 3.  However, for $k$, since its ages were 4 and 4, after
join, the age remains 4.
We define the join operator in this way because our goal is to conduct
a \emph{Must-Hit} analysis: we know that a variable $v\in V$ is
definitely in the cache \emph{only if} $v$ is in the cache according
to both states before the join, i.e., $Age(v) \leq N$ and
$Age'(v) \leq N$.

Formally, given two states $S = \langle
Age(v_1),\ldots,Age(v_n) \rangle$ and $S' = \langle
Age(v'_1),\dots,Age(v'_n) \rangle$, we define $S'' = S \sqcup S'$ as
follows: $S'' = \langle max(Age(v_1),Age(v'_1)), \ldots,
max(Age(v_n),Age(v'_n)) \rangle$.

\section{Modeling the Speculative Execution}
\label{sec:approach}

In this section, we lift the baseline abstract interpretation
algorithm so that it can soundly model speculative execution.

\subsection{Augmented CFG with Virtual Control Flow}

Given the CFG of a program, we first augment it by adding special
nodes and edges, to model all possible control flows produced by
speculative executions.  These implicit control flows, which will be
made explicit in our augmented CFG, are called the \emph{virtual
control flows}.

A virtual control flow occurs at every \emph{if-else} statement where
the branching condition depends on some variables stored in memory.
In a normal execution, a branch guarded by a condition $(c)$ is
explored only when $c$ is satisfied.  However, under speculative
execution, the branch will be explored (speculatively) by our
algorithm even if $c$ is unsatisfiable.  Furthermore, upon
mis-prediction, the rollback will re-direct the control to the other
branch.

To model all these behaviors, we add the following special nodes and
edges to the CFG for every branch that may be explored speculatively:
\begin{itemize}
\item 
$vn_{start}$, which is a special CFG node that denotes the start of a
virtual control flow;
\item
$vn_{stop}$, which is a special CFG node that denotes the end of a
virtual control flow.
%\item
%$ve$, which is a special CFG edge denoting a control flow under
%speculative execution.  
\end{itemize}
The edges connecting such nodes, which represent the virtual control
flows, fall into five categories: (1) $n$--$vn_{start}$; (2)
$vn_{start}$--$n$; (3) $n$--$n$; (4) $n$--$vn_{stop}$, and (5)
$vn_{stop}$--$n$, where $n$ is a normal CFG node.

The edge $n$--$vn_{start}$ represents the start of a speculative
execution: it feeds the state $S[n]$ to $vn_{start}$, which in turn
generates a speculative state $SS[vn_{start}]=S[n]$.
Then, the newly created speculative state is propagated through the
edge $vn_{start}$--$n$.  Next, it is propagated through the edges
$n$--$n$ and $n$--$vn_{stop}$ until reaching $vn_{stop}$--$n$.
The special node $vn_{stop}$ converts the speculative state
$SS[vn_{stop}]$ back to the normal state $S[n] = SS[vn_{stop}]$.
Afterward, the state is joined with other states from the normal
execution.

\begin{figure}
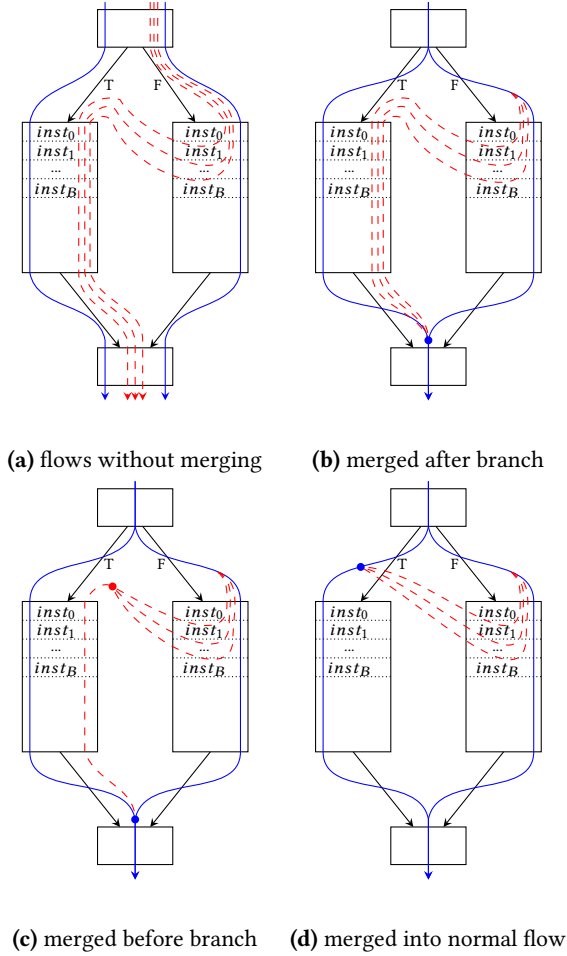
%[htb!]
	\begin{subfigure}[t]{.45\linewidth}
		\centering\include{fig/trace1}
		\caption{flows without merging}
		\label{fig:trace1}
	\end{subfigure}
	\begin{subfigure}[t]{.45\linewidth}
		\centering\include{fig/trace2}
		\caption{merged after branch}
		\label{fig:trace2}
	\end{subfigure}
	\begin{subfigure}[t]{.45\linewidth}
		\centering\include{fig/trace3}
		\caption{merged before branch}
		\label{fig:trace3}
	\end{subfigure}
	\begin{subfigure}[t]{.45\linewidth}
		\centering\include{fig/trace4}
		\caption{merged into normal flow}
		\label{fig:trace4}
	\end{subfigure}
	\caption{Strategies for merging speculative control flows.}
	\label{fig:merge}
\vspace{-2.5pt}
\end{figure}

%Although every instruction in the program may be speculatively
%executed, only those that can be rolled back due to mis-prediction
%will cause side effects on the cache. Thus, we will generate the
%virtual control flows by focusing on these instructions.
%
One way to add the special nodes and edges is illustrated in
Figure~\ref{fig:trace1}.  Specifically, for each \emph{if-else}
statement, we add virtual control flow edges from instructions in one
of the branch to the entry node of the other branch under the same
branching condition.

Here, the blue solid lines represent normal executions, whereas the
red dashed lines represent virtual control flows associated with
speculative executions of the \emph{else}-branch.  
\shepherd{Virtual control flows associated with the 
\emph{then}-branch are similar, but omitted in the figure for clarity.}
The reason why there are more than one dashed lines is because the
roll-back point (i.e., location where roll-back occurs) is
non-deterministic; to be conservative, we assume it may occur at any
moment within the maximum speculation depth.

In practice, the \emph{speculation depth} is platform-dependent and
bounded by a few factors~\cite{PierceM94, FilhoF97}, e.g., the size of
the reorder buffer; the maximum number of unresolved branches that the
CPU can handle before it stalls; whether there are division-by-zero or
floating-point errors in the program; and the number of clock cycles
taken to access memory and resolve a branching condition.
For simplicity, for example, we assume that the maximum speculative
execution depth is provided by the user.
In Figure~\ref{fig:trace1}, we assume that $inst_B$ is the
boundary within which roll-back occurs.

\subsection{Merging the Speculative Flows}
\label{sec:merge}

Since we use abstract interpretation to over-approximate the cache
states, multiple executions must be merged to reduce the computational
overhead.  In the baseline algorithm, for example, states from two
different paths are joined whenever the program paths are merged in
the CFG.  In the speculative analysis, we also need to decide when to
join the normal and the speculative states.

Figure~\ref{fig:merge} shows three merging strategies in addition to
the original \emph{no-merging} strategy in Figure~\ref{fig:trace1}.
Consider Figure~\ref{fig:trace2}, for example, since the executions
before the branch entry node are identical, they are merged without
losing accuracy; in addition, the speculative executions are merged
right before the exit point of the other branch.  Recall that the join
operator ($\sqcup$) used to handle merging is over-approximated, we
know that the strategy outlined in Figure~\ref{fig:trace2} is a sound
over-approximation of Figure~\ref{fig:trace1}.

To over-approximate even more, consider Figure~\ref{fig:trace3}, which
merges all speculative states of the \emph{else}-branch before
reaching the \emph{then}-branch.  However, the merged speculative
state is propagated through the \emph{then}-branch before it is merged
with the normal state. 
In contrast, Figure~\ref{fig:trace4} is a more aggressive
over-approximation, which merges the speculative states with the
non-speculative state at the entry node of the \emph{then}-branch.

Regardless of the merging strategy, however, our method ensures that
the result is a sound over-approximation.  
\shepherd{
Since every time state merging occurs, it may lose information, in
general, the later that merging occurs, the more accurate the result
is, but there is no guarantee. Furthermore, late merging may lead to a
more expensive analysis.  }
Our experimental comparisons of these four strategies show that the
one outlined in Figure~\ref{fig:trace3} is the best: it not only
obtains significantly more accurate results than the one in
Figure~\ref{fig:trace4}, but also runs almost equally fast.
Therefore, we have settled down on this strategy: we call
it \emph{Just-in-Time} merging.

\subsection{Just-in-Time Merging: An Example}

\begin{figure}
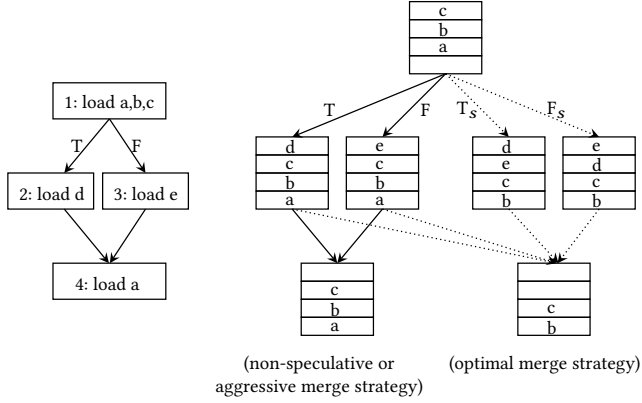

	\include{fig/merge}
	\caption{Cache state with different merge points.}
	\label{fig:mergeCache}
\end{figure}

Consider the CFG of a branch shown on the left-hand side of
Figure~\ref{fig:mergeCache}, where each basic block refers to a
variable (from $a$ to $e$).  The initial cache state, at the top of
the figure on the right-hand side, is the state after executing the
first basic block, where variables $a$, $b$ and $c$ are loaded into
the cache.  Here, the solid arrows represent the normal execution,
where either $d$ or $e$ is mapped to the youngest cache line.  Since
we are concerned with a \emph{Must-Hit} analysis, after merging at
basic block 4, only $a$, $b$ and $c$ are left in the cache.

Under speculative execution, we may execute the \emph{else}-branch
before rolling back to execute the \emph{then}-branch. If we choose to
merge the speculative state right after the rollback, the merging
would be between $d$, $c$, $b$ and $a$ on the one hand, and $e$, $d$,
$c$ and $b$ on the other hand.  The merged state will not contain $e$
anymore, thus losing the important information of speculative
execution.

However, if we propagate the speculative state computed from
the \emph{else}-branch through the \emph{then}-branch and then merge
with the non-speculative state, the cache state at basic block 4 will
be more accurate.
As shown by the dotted arrow \textit{T$_s$}, variable $e$ is loaded to
the cache before $d$ is loaded to the cache; similarly,
for \textit{F$_s$}, variable $d$ is loaded before $e$ is
loaded. Finally, when the four states are merged, the result is that
only $c$ and $b$ are guaranteed to result in cache hits.
Thus, the cache state on the bottom-right of
Figure~\ref{fig:mergeCache}, which corresponds to
\emph{Just-in-Time} merging  illustrated in
Figure~\ref{fig:trace3}, captures the side effect of speculative
execution.

\section{Generalization and Optimization}
\label{sec:optimization}

In this section, we present the generalized algorithm before
discussing several optimizations, which help increase accuracy as well
as decrease runtime overhead.
%\subsection{Abstract Interpretation under Speculation}

Algorithm~\ref{alg:pseudo} shows the static analysis procedure that is
sound under speculative execution.  Given the original CFG of a
program, it first constructs an augmented CFG by adding the virtual
control flows.  Then, it initializes the abstract states for each
program location $n$, including both the default state, denoted
$S[n]$, and the speculative state, denoted $SS[n]$.  Next, it starts
the fixed-point computation using a worklist based procedure that is
similar to that of Algorithm~\ref{alg:baseline}.

\begin{algorithm}
	\caption{Abstract interpretation under speculation.}
	\label{alg:pseudo}
	\footnotesize
	\begin{algorithmic}[1]
		\State \textcolor{darkblue}{$\mathit{VCFG} \gets$ \Call{AugmentCFGwithVirtualControlFlow}{$\mathit{CFG}$}}
                \State Initialize $S[n]$ to $\top$ if $n \in$ \Call{Entry}{$\mathit{VCFG}$}, else to $\bot$ 
                \State \textcolor{darkblue}{Initialize $SS[n]$ to $\bot$ for all $n\in \mathit{VCFG}$}

		\State $WL \gets$ \Call{Entry}{$\mathit{VCFG}$}
		\While{$\exists n \in WL$ }
		\State $WL \gets WL \setminus \{n\}$
                
                \If { $n$ is a normal CFG node }
		  \State $s' \gets $ \Call{Transfer}{ $S[n]$, $n$}
		  \State \textcolor{darkblue}{$ss' \gets $ \Call{Transfer}{ $SS[n]$, $n$}}
                \Else
                  \State \textcolor{darkblue}{Set $ss'$ to $S[n]$  if $n$ is a special $n_{start}$ node, else to $\bot$}
                  \State \textcolor{darkblue}{Set $s'$ to $SS[n]$  if $n$ is a special $n_{stop}$ node, else to $\bot$}
                \EndIf

		\ForEach{$n' \in $ \Call{Successors}{$\mathit{VCFG}$, $n$}}
		  \If{$s' \not \sqsubseteq S[n']$ \textcolor{darkblue}{or $ss' \not \sqsubseteq SS[n']$} }
		    \State  $ S[n'] \gets$ $S[n'] \sqcup s'$
		    \State  \textcolor{darkblue}{$ SS[n'] \gets$ $SS(n') \sqcup ss'$}
		    \State  $WL \gets WL \cup \{n'\}$
                  \EndIf
                \EndFor

		\EndWhile
	\end{algorithmic}
\end{algorithm}

However, when the special CFG node $vn_{start}$ is encountered (Line
11), the default state $S[n]$, which is from the incoming edge, is
used to create a speculative state $ss'\gets S[n]$; this is to model
the side effects caused by the failed speculative execution upon
rollback.  
From then on, both the default state $S[n]$ and the speculative state
$SS[n]$ will be propagated through  subsequent nodes in the VCFG;
at each node $n$, the transfer function has to be applied to both of
them (Lines~8-9).  This continues until the other special node
$vn_{stop}$ is encountered, which transforms the speculative state
$SS[n]$ back to $s'$ (Line~12) before $s'$ is merged into the normal
flow.

\ignore{%%%% original verison of the speculative algorithm %%%%

\begin{algorithm}[t!]
	\caption{Speculative Cache Analysis}
	\label{alg:pseudo}
	\footnotesize
	\begin{algorithmic}[1]
		\Function {SpecuCacheAnalysisPass}{$CFG$}
		\State $Cs(bb), SCs(bb)$ initialized to $\top$ if $bb \in$ \Call{Entry}{$CFG$}, else to $\bot$ 
		\State $CFG \gets$ \Call{GenVCFG}{$CFG$}
		\State $WL \gets$ \Call{Entry}{$CFG$}
		\While{$\exists$ $bb \in WL$ }
		\State $WL \gets WL \setminus \{bb\}$
		\If{$bb$ is \textcolor{darkblue}{darkblue} node}
		\State $s' \gets $ \Call{T}{$Cs(bb)$, $bb$}
		\Else
		\State $s' \gets $ \Call{T}{$SCs(bb)$, $bb$}
		\EndIf
%		\State $CFG \gets$ \Call{DynBoundVCFG}{$CFG$, $bb$, $s'$}
		\ForAll{$bb' \in $ \Call{Succ}{$CFG$, $bb$}}
		\If{$bb'$ is \textcolor{darkblue}{darkblue} node $\&$ $s' \not \sqsubseteq Cs(bb')$}
		\State $ Cs(bb') \gets$ \Call{Widen}{$Cs(bb')$, $s'$}
		\State $WL \gets WL \cup \{bb'\}$
		\ElsIf{$bb'$ is \textcolor{red}{red} node $\&$ $s' \not \sqsubseteq SCs(bb')$}
		\State $ SCs(bb') \gets$ \Call{Widen}{$SCs(bb')$, $s'$}
		\State $WL \gets WL \cup \{bb'\}$
		\EndIf
		\EndFor
		\EndWhile
		\EndFunction
	\end{algorithmic}
\end{algorithm}

}%\ignore{%%%% original version of the speculative algorithm %%%%}

\subsection{The Running Example}

\begin{figure}
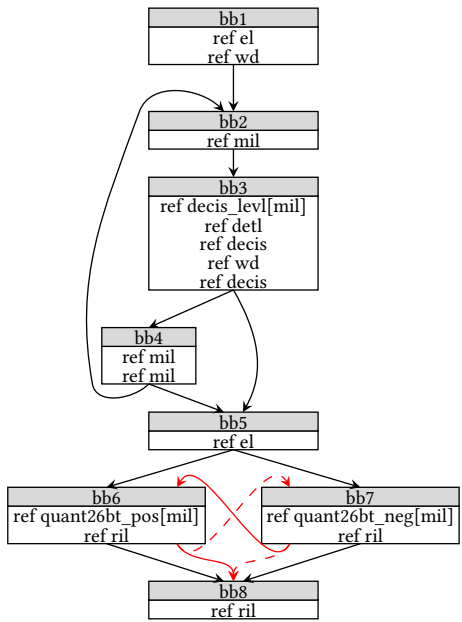

\vspace{3ex}
\centering
\include{fig/realex}
\caption{Augmented CFG with virtual control flows.}
\label{fig:realexcfg}
	%note: change ref to read & store, make it more clear
\end{figure}

To illustrate how the algorithm works, consider the example program in
Figure~\ref{fig:realEx}, which is a real-time DSP program written in
C~\cite{WCET2010Benchmarks}. The corresponding CFG is shown in
Figure~\ref{fig:realexcfg}, where the red (solid and dashed) edges
represent the two virtual control flows.

\begin{figure}
\vspace{1ex}

\hspace{4ex}
\begin{minipage}{.95\linewidth}
\begin{lstlisting}[numbers=left,language=C]
/* table is 31-byte long to make quantl look-up 
easier, last entry is for mil=30 when wd is max */
int quant26bt_pos[31] = { 61,60,59,58,57,56,55,54,
  53,52,51,50,49,48,47,46,45,44,43,42,41,40,39,
  38,37,36,35,34,33,32,32 };
/* table is 31-byte long to make quantl look-up
easier, last entry is for mil=30 when wd is max */
int quant26bt_neg[31] = { 63,62,31,30,29,28,27,26,
  25,24,23,22,21,20,19,18,17,16,15,14,13,12,11,10,
  9,8,7,6,5,4,4 };
/* decision levels - pre-multiplied by 8 */
int decis_levl[30] = { 280,576,880,1200,1520,1864,
  2208,2584,2960,3376,3784,4240,4696,5200,5712,
  6288,6864,7520,8184,8968,9752,10712,11664,12896,
  14120,15840,17560,20456,23352,32767 };

int quantl(int el,int detl) {
	int ril,mil;
	long int wd,decis;
	/* abs of difference signal */
	wd = my_abs(el);
	/* mil based on decision levels and detl gain */
	for(mil = 0 ; mil < 30 ; mil++) {
	  decis = (decis_levl[mil]*(long)detl) >> 15L;
	  if(wd <= decis) break;
	}
	/*if mil=30, wd is less than all decision levels*/
	if(el >= 0) ril = quant26bt_pos[mil];
	else ril = quant26bt_neg[mil];
	return(ril);
}\end{lstlisting}
	\end{minipage}
	\caption{Code snippet from a real-time DSP program~\cite{WCET2010Benchmarks}.}
	\label{fig:realEx}
\end{figure}

\paragraph{Result from Non-speculative Executions}

Table~\ref{tbl:cacheTrace} shows the cache state computed for each
location (basic block) based on only normal executions (black edges in
Figure~\ref{fig:realexcfg}); this is analogous to running the baseline
procedure in Algorithm~\ref{alg:baseline}.  In Column~2, the variables
are arranged according to their ages: the younger variable appears on
the left.

\begin{table}[t!]
	\caption{Cache states during the fixed-point computation.}
	\label{tbl:cacheTrace}
	\centering
	\scalebox{.7}{
		\begin{tabular}{|c|l|}
			\hline
			BBlk  &Cache State  \\ 
			\hline\hline
			0 & \{\hspace{5pt}\} \\
			1 & \{wd, el\} \\
			2 & \{mil, wd, el\} \\
			3 & \{decis, wd, detl, decis\_lev[1*], mil, el\}\\ 
                        4 & \{mil,decis, wd, detl, decis\_lev[1*], el\}\\   
%			5 & \{el, decis, wd, detl, decis\_lev[1*], mil\}\\   
			2 & \{mil,decis, wd, detl, decis\_lev[1*], el\}\\   
%			6 & \{ril, quant26bt\_pos[1*], el, decis, wd, detl, decis\_lev[1*], mil\}\\   
%			7 & \{ril, quant26bt\_neg[1*], el, decis, wd, detl, decis\_lev[1*], mil\}\\   
			3 & \{decis, wd,detl, decis\_lev[2*], mil, decis\_lev[1*], el\}\\ 
%			8 & \{ril,$\emptyset$ , el, decis, wd, detl, decis\_lev[1*], mil\}\\   
			4 & \{mil, decis, wd,detl, decis\_lev[2*], decis\_lev[1*], el\} \\
			2 & \{mil, decis, wd,detl, decis\_lev[2*], decis\_lev[1*], el\} \\
			5 & \{el ,decis, wd,detl, decis\_lev[2*], mil, decis\_lev[1*]\}\\   
			\rowcolor{lightblue}
			6 & \{ril, quant26bt\_pos[1*], el ,decis, wd,detl, decis\_lev[2*], mil, decis\_lev[1*]\}\\
			\rowcolor{lightblue}
			7 & \{ril, quant26bt\_neg[1*], el ,decis, wd,detl, decis\_lev[2*], mil, decis\_lev[1*]\}\\
			\rowcolor{lightgray}
%			3 & \{decis, wd,detl, decis\_lev[2*], mil, decis\_lev[1*], el\}\\ 
			8 & \{ril, $\emptyset$ , el ,decis, wd,detl, decis\_lev[2*], mil, decis\_lev[1*]\}\\   
			\hline
		\end{tabular}
	}
\end{table}

Initially, the cache is empty. From basic block 1 to 5, we apply the
transfer functions: \texttt{decis\_lev} takes two cache lines, but
since we do not unwind the loop, we do not know its index
statically. Thus, we nondeterministically pick one for the first
time, \texttt{decis\_lev[1*]}.
Following the back edge from basic block 4, when \texttt{decis\_lev}
is accessed again, we conservatively choose the second cache line
for \texttt{decis\_lev[2*]} to ensure that the cache state remains an
over-approximation.
Our analysis iterates through the loop three times before it reaches
a fixed-point (light gray row) and terminates.

\paragraph{Result from Speculative Executions}

%To demonstrate the differences in the computed cache state, we compare
%the results obtained by Algorithm~\ref{alg:pseudo} to those obtained
%by Algorithm~\ref{alg:baseline}.  This time, the red (solid and
%dashed) edges in Figure~\ref{fig:realexcfg}, which represent the
%virtual control flows, are also considered . 
%
%That is, at Line 28 of the program in Figure~\ref{fig:realEx}, the
%branch condition \textit{(el>=0)} is controlled by input, which may be
%manipulated to mis-train the branch predictor. Here, we assume that
%arrays \textit{quant26bt\_pos} and \textit{quant26bt\_neg} map to
%different cache lines. 

Table~\ref{tbl:cacheTraceSpecu} shows the cache state computed under
speculative execution.
For clarity, we only focus on the cache states relevant to the
speculative executions starting from basic block 5.  We use two
different colors, blue and red, to differentiate the cache states
computed from non-speculative (blue) and speculative (red) executions.
By considering speculative executions, it is possible for us to access
both \texttt{quant26bt\_pos} and \texttt{quant26bt\_neg} in a single
execution.

\paragraph{Execution Time Estimation}

The last row of Table~\ref{tbl:cacheTraceSpecu}, which differs from
the last row of Table~\ref{tbl:cacheTrace}, shows that most of the
program variables have older ages than before.
This is dangerous because, if the cache is only large enough to hold
the first eight variables, there will be an additional cache miss,
which may force the program to miss its deadline.

\begin{table}[t!]
	\caption{Cache states during speculative execution.}
	\label{tbl:cacheTraceSpecu}
	\centering
	\scalebox{.7}{
		\begin{tabular}{|c|p{1.2\linewidth}|}
			\hline
			BBlk  &Cache State  \\ 
			\hline\hline
			\rowcolor{white}
			... & ...\\
%			\rowcolor{lightblue}
%			3 & \{decis, wd,detl, decis\_lev[2*], mil, decis\_lev[1*], el\}\\ 
%			\rowcolor{lightblue}
%			4 & \{mil, decis, wd,detl, decis\_lev[2*], decis\_lev[1*], el\} \\
%			\rowcolor{lightblue}
%			2 & \{mil, decis, wd,detl, decis\_lev[2*], decis\_lev[1*], el\} \\
			\rowcolor{white}
			5 & \{el ,decis, wd,detl, decis\_lev[2*], mil, decis\_lev[1*]\}\\   
			\rowcolor{lightblue}
			6 & \{ril, quant26bt\_pos[1*], el ,decis, wd,detl, decis\_lev[2*], mil, decis\_lev[1*]\}\\
			\rowcolor{lightblue}
			7 & \{ril, quant26bt\_neg[1*], el ,decis, wd,detl, decis\_lev[2*], mil, decis\_lev[1*]\}\\
			\rowcolor{lightred}
			6 & \{ril, quant26bt\_pos[1*], el ,decis, wd,detl, decis\_lev[2*], mil, decis\_lev[1*]\}\\
			\rowcolor{lightred}
			7 & \{ril, quant26bt\_neg[1*], el ,decis, wd,detl, decis\_lev[2*], mil, decis\_lev[1*]\}\\
			\rowcolor{lightred}
			7 & \{ril, quant26bt\_neg[1*], quant26bt\_pos[1*], el ,decis, wd,detl, decis\_lev[2*], mil, decis\_lev[1*]\}\\
			\rowcolor{lightred}
			6 & \{ril, quant26bt\_pos[1*], quant26bt\_neg[1*], el ,decis, wd,detl, decis\_lev[2*], mil, decis\_lev[1*]\}\\
			\rowcolor{lightred}
			8 & \{ril, $\emptyset$ , el ,decis, wd,detl, decis\_lev[2*], mil, decis\_lev[1*]\}\\
			\rowcolor{lightgray}
			8 & \{ril, $\emptyset$ , $\emptyset$, el ,decis, wd,detl, decis\_lev[2*], mil, decis\_lev[1*]\}\\   
			\hline
		\end{tabular}
	}
\end{table}

\paragraph{Side Channel Detection}

The additional cache miss may also lead to side-channel leaks.
Figure~\ref{fig:cache_harness} shows a client program that uses the
program in Figure~\ref{fig:realEx}.  The application first accepts
some input from the user, then processes it using \textit{quantl} as a
subroutine, and finally encrypts the result using a cipher such as
AES.
Before calling \textit{quantl}, a look-up table named \textit{sbox} is
loaded; the lookup table will be used by the cipher while it encrypts
the data, during which time a secret \emph{key} is used as the index
to access \textit{sbox}.

\begin{figure}
	\centering
	\begin{minipage}{.9\linewidth}
		\begin{lstlisting}[numbers=left,language=C]
#define BUF_SIZE 1024*16
const uint8_t sbox[256] = { 0x63, 0x7c, 0x77, 
0x7b, 0xf2, 0x6b, 0x6f, 0xc5, 0x30, 0x01, 0x67, 
0x2b, 0xfe, 0xd7, 0xab, 0x76, ... };
int main()
{
 uint32_t inBuf[BUF_SIZE];
 int el, delt, tmp;
 for(int i=0; i< 256; i++) // preload sbox
  tmp = sbox[i]; 	
 for(int i=0; i< BUF_SIZE; i++) // read inBuf
  tmp = inBuf[i]; 
 tmp = quantl(el, delt);	
 AES_encode(inBuf);
}\end{lstlisting}
	\end{minipage}
	\caption{The client code that leads to side-channel leaks.}
	\label{fig:cache_harness}
\end{figure}

By controlling the input size, a malicious user can force part of
the \textit{sbox} to be evicted from the cache. As a result, for
some \emph{key} values, accessing \textit{sbox} results in a cache
hit, but for other \emph{key} values, it results in a cache miss.
Although timing side channels have been investigated 
before~\cite{DoychevFKMR13,BartheKMO14,WuGSW18,GuoWW18}, these prior works never considered speculative execution.  Our contribution, in this 
context, is to show that even if a program is \emph{leak-free} under
normal execution, it may still be leaky under speculative execution.

\subsection{Dynamically Bounding Speculation Depth}

Although the maximum number of speculatively executed instructions is
used to construct the augmented CFG, in practice, the number of
speculatively executed instructions can be smaller.
For example, when all variables needed to resolve a branching
condition are in the cache, speculative execution may be
shortened.  Since our cache analysis aims to decide whether a variable
access is a \emph{must-hit}, as the analysis continues it may report
more \textit{must-hit} variables, which can be used to bound the
speculation depths of other branches.

Thus, we propose an optimization that leverages
the \emph{must-hit} variables to dynamically remove virtual control
flows that are deemed redundant.
Toward this end, we maintain two predefined bounds for each
speculative execution, $b_h$ and $b_m$, which correspond to the
branching condition being a cache hit or miss. (Since $b_h$ and $b_m$
are platform-dependent, they are set based on input from the user.)
By default, we use $b_m$ as the bound; but as soon as the branching
condition is proved to be a must-hit, we switch the bound to $b_h$.

This optimization not only decreases the computational overhead, i.e.,
by reducing the number of edges in the VCFG, but also increases the
accuracy since it results in a potentially tighter over-approximation.
In the extreme case where $b_h=0$, for example, switching to $b_h$
means avoiding speculative execution all together, which can avoid 
many bogus behaviors.

\shepherd{
While our focus here is on exploiting changes to the speculation depth
due to cache misses, the proposed technique may be extended to exploit
other sources of changes, e.g., execution units being busy, or
division taking a longer time based on the operands. }
%
%To make it easier to understand, let's assume
%$b_h$ to be zero, which means no instruction will be speculatively
%executed if branch condition is in the cache. By correctly identifying
%such cases, our analysis can exclude more redundant behaviors to
%reduce the number of false cache misses.  In such a case, dynamic
%boundary could reduce false misses reported by the analysis.

\vspace{-2ex}
\subsection{Handling the Merges and Loops }

The algorithm presented so far uses the join operator ($\sqcup$) to
over-approximate the union of two abstract states.  However, in the
presence of loops, it may have limitations: (1) the resulting state
may not be accurate enough, and (2) it may take a long time (or
forever) to reach a fixed point.

Thus, we add a widening operator~\cite{Cousot1978} to the standard
join operation $s[n'] \sqcup s'$; that is, we use $(s[n'] \sqcup
s') \nabla s'$ instead of $s[n'] \sqcup s'$.
The idea behind widening ($\nabla$) is simple: first, we identify
the \emph{direction of growth} from the state $s'$ to the state
$(s[n'] \sqcup s')$; then, we over-approximate 
$(s[n'] \sqcup s')$ in such a way that it maximizes the progress along
the \emph{direction of growth}.
In the interval domain, for example, if the previous state is $s' =
0\leq x \leq 3$ and the current state is $s = 0\leq x \leq 5$, the
result of widening would be $s \nabla s' = 0 \leq x \leq +\infty$.
\shepherd{To achieve better accuracy, 
loops with fixed iteration number will be fully unrolled; only
unresolved loops will be widened.  }

\begin{figure}
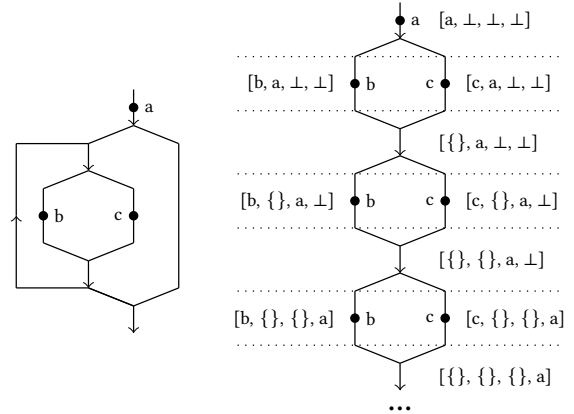

\centering
\begin{minipage}{.15\textwidth}
\include{fig/widening}
\end{minipage}
\begin{minipage}{.3\textwidth}
\include{fig/widening1_v2}
\end{minipage}
\vspace{-5ex}
\caption{Example program for the widening operator.}
\label{fig:widening}
\end{figure}

Figure~\ref{fig:widening} shows another loop-related problem.  First,
variable \texttt{a} is loaded into the cache. Then, inside the loop,
every time the branch is executed, $Age(a)$ increases by 1. After the
join, however, neither \texttt{b} nor
\texttt{c} will be in the cache. 
Thus, eventually, $a$ is evicted from the cache as well.  This is not
accurate because, during the actual execution, $a$ will never be
evicted.
With a refined join operator, we will be able to avoid this problem.

%In Figure~\ref{fig:widening}, we use array \texttt{b[k]} instead
%of \texttt{b} to illustrate the slow convergence of loop, because in
%reality the program can gradually load array \texttt{b[]} to
%evict \texttt{a} out of cache. However, when we use single
%variable \texttt{b} instead of array, the analysis will still
%age \texttt{a} at each loop iteration, since both \texttt{b}
%and \texttt{c} will no longer in the cache at the branch merge at the
%end of loop. Obviously, \texttt{a} should always stay in the cache no
%matter how many iteration the loop runs, as there are only 3 variables
%in the program. This interesting observation is due to the information
%loss at state merging. To improve the accuracy of our analysis in such
%situations, we borrow the idea from Definitely Unknown
%Analysis~\cite{touzeau2017} to improve the accuracy and
%also accelerate the termination of loops.

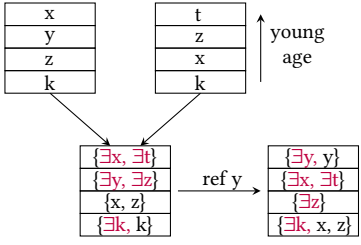
\begin{figure}%[htb!]
\centering
\scalebox{1}{\input{fig/newcachejoin_v2}}
\caption{Transfer function  with shadow variables.}
\label{fig:newcachejoin}
\end{figure}

We refine the join operator ($\sqcup$) by adding extra information
into the cache state.
Similar to Touzeau et al.~\cite{touzeau2017,TouzeauMMR19}, we
introduce a shadow variable $\exists v$ for each $v \in V$ to
represent \emph{the youngest cache line} that may hold $v$ along some
program path.
Whenever two states are merged and $v$ appears in only one of the two
states, $\exists v$ will remain in the merged state while the
variable $v$ will not.
Figure~\ref{fig:newcachejoin} shows an example, where both
$\exists$\texttt{y} and $\exists$\texttt{t} appear in the merged
state. That is, there exists some path in which \texttt{y} (or
\texttt{t}) is cached.

We also revise the transfer function, by using the number of shadow
variables \emph{younger than or equal to $u$} to refine the rule for
aging the regular variable $u$~\cite{NagarS12}.
In Figure~\ref{fig:newcachejoin}, for example, if \texttt{y} is
accessed on the merged state, the ages of $x,z$ increase by 1 and $k$
is evicted from the cache.
Details of this optimization can be found in
Appendix~\ref{app:modified}.

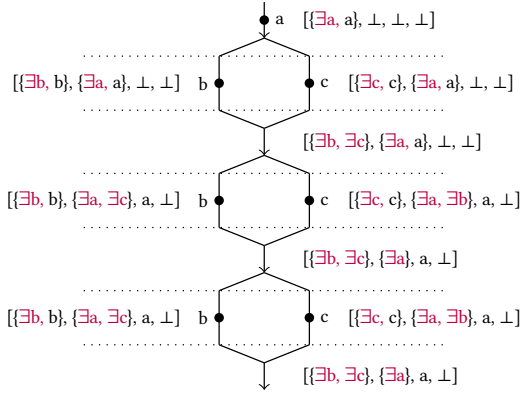
\begin{figure}%[htb!]
\centering
\scalebox{1.2}{\input{fig/exist_hit_v2}}
% \scalebox{1}{\input{fig/widening1}}
\caption{The refined join using shadow variables.}
\label{fig:existhit}
\end{figure}

For simplicity, we unroll the loop for three times and illustrate the
sequence of memory accesses in Figure~\ref{fig:existhit}. The abstract
cache states are listed on both sides at each memory access and merge
point. With the shadow variables, our cache states are able to reach
the fixed-points after only three iterations and avoid evicting
\texttt{a}.

\subsection{Handling Multiple Speculative Executions}

Finally, we extend our algorithm so it can independently propagate the
speculative states through the virtual control flows, without
interfering each other, even if one branching statement is embedded
inside another branching statement.  
%Although such scenarios are
%soundly handled by the algorithms presented so far, i.e., by eagerly
%merging these speculative states, the accuracy is not as high as
%propagating them independently.

Algorithm~\ref{alg:pseudo2} shows the procedure, which computes, for
each node $n$ in the augmented CFG, a set of states of the form
$SS[n][c]$, one for each speculative execution.  Let
$C=\{1,\ldots,k\}$ be the set of all branches in the program that can
be speculatively executed; each $1\leq i\leq k$ is the index of a
branch in this set.  We call $c=i$ the color of the $i$-th speculative
execution.
While constructing the VCFG, for each $c\in C$, we add a separate set
of virtual control-flow edges and nodes, with the color
$c$.

\begin{algorithm}[t!]
	\caption{Analysis under a set of speculative executions.}
	\label{alg:pseudo2}
	\footnotesize
	\begin{algorithmic}[1]
		\State \textcolor{darkblue}{($\mathit{VCFG},C) \gets$ \Call{AugmentCFGwithVirtualControlFlow}{$\mathit{CFG}$}}
                \State Initialize $S[n]$ to $\top$ if $n \in$ \Call{Entry}{$\mathit{VCFG}$}, else to $\bot$ 
                \State \textcolor{darkblue}{Initialize $SS[n][c]$ to $\bot$ for all $n\in \mathit{VCFG}$ and for all color $c\in C$}

		\State $WL \gets$ \Call{Entry}{$\mathit{VCFG}$}
		\While{$\exists n \in WL$ }
		\State $WL \gets WL \setminus \{n\}$
                
                \If { $n$ is a normal CFG node }
		  \State $s' \gets $ \Call{Transfer}{ $S[n]$, $n$}
		  \State \textcolor{darkblue}{$ss'[c] \gets $ \Call{Transfer}{ $SS[n][c]$, $n$} for all color $c\in C$}
                \Else
                  \State \textcolor{darkblue}{Set $s'$ to $SS[n][c]$  if $n$ is node $n_{start}$ of color $c$, else to $\bot$}
                  \State \textcolor{darkblue}{Set $ss'[c]$ to $S[n]$  if $n$ is node $n_{stop}$ of color $c$, else to $\bot$}
                \EndIf

		\ForEach{$n' \in $ \Call{Successors}{$\mathit{VCFG}$, $n$}}
		  \If{$s' \not \sqsubseteq S[n']$ \textcolor{darkblue}{or $\exists c: ss'[c] \not \sqsubseteq SS[n'][c]$} }
		    \State  $ S[n'] \gets$ $S[n'] \sqcup s'$
		    \State  \textcolor{darkblue}{$ SS[n'][c] \gets$ $SS(n') \sqcup ss'[c]$ for all color $c\in C$}
		    \State  $WL \gets WL \cup \{n'\}$
                  \EndIf
                \EndFor

		\EndWhile
	\end{algorithmic}
\end{algorithm}

During the fixed-point computation, instead of applying the transfer
function once to generate a speculative state $ss'$, the procedure
applies the transfer function $|C|$ times, to generate a vector of
speculative states $ss'[c]$, one for each speculative execution with
color $c$.
As such, every speculative execution (of color $c\in C$) is handled
separately until the corresponding node $n_{stop}$ (of the same color
$c$) is encountered, in which case the speculative state $SS[n][c]$ is
transformed back to a non-speculative state $s'$.

\shepherd{
There are alternative ways of presenting the analysis procedure in
Algorithm~\ref{alg:pseudo2}, for example, by using the trace
partitioning framework developed by Mauborgne and
Rival~\cite{MauborgneR05}.  Also note that, for ease of comprehension, we choose to
split the speculative states from the normal states. However, the two
types of states may be treated uniformly and processed using a
generalized worklist-based algorithm.  Assume that the worklist-based
algorithm is smart enough, the special merge nodes created for virtual
control flows can be viewed as merely optimization hints. }

%\newpage
\section{Experiments}
\label{sec:experiment}

We have implemented our method in LLVM~\cite{LLVM} and experimentally
compared it with a state-of-the-art, non-speculative static cache analysis
technique~\cite{WuGSW18}.  In our experiments, we used a set-associative 
cache with the LRU replacement policy, 512 cache lines,
and 64 bytes per line.  The speculative execution depths, following a
cache hit and a cache miss, are set to 20 and 200 instructions, respectively.
These bounds were derived from our analysis of the pipelined execution
traces produced by GEM5~\cite{GEM5}, a state-of-the-art
micro-architecture simulator, with \emph{O3CPU}, which is a detailed
out-of-order CPU model based on the Alpha 21264 processor.

Our experiments were designed to answer three questions:
%\begin{itemize}
%\item 
(1) Is our method more accurate in detecting cache misses than
the existing method, which does not consider speculative execution?
%\item 
(2) Is our method fast enough for practical use?
%\item 
(3) Are the optimizations proposed in Section~\ref{sec:optimization}
effective in reducing overhead and increasing accuracy?
%\end{itemize}

\subsection{Benchmarks}

Tables~\ref{tbl:bench1} and~\ref{tbl:bench2} show the statistics of
our benchmarks, collected from various sources including the
\emph{Malardalen} real-time software benchmark~\cite{WCET2010Benchmarks}, a
commercially representative embedded software suite named
\emph{MiBench}~\cite{guthaus2001mibench}, a high performance patch for SSH
(\emph{hpn-ssh})~\cite{hpn-ssh}, a cryptographic toolkit
named \emph{LibTomCrypt}~\cite{LibTomCrypt}, the \emph{openssh} source
code~\cite{openssh}, and a Linux kernel for \emph{tegra}~\cite{Tesla}
used on Tesla automobiles.
These benchmarks are divided into two sets: execution time estimation
and side channel detection.
The benchmarks for execution time estimation (Table~\ref{tbl:bench1})
are used as is, whereas the benchmarks for side channel detection
(Table~\ref{tbl:bench2}) are used together with a client program that
we wrote, to invoke the benchmark program in a way similar to
Figure~\ref{fig:cache_harness}.

\begin{table}[t!]
	\caption{Execution time estimation: benchmark statistics.}
	\label{tbl:bench1}
	\centering
	\scalebox{.7}{
		\begin{tabular}{|llp{.8\linewidth}r|}
			\hline
			Name  &Source &Description &Loc   \\ 
			\hline\hline
			adpcm          &WCET@mdh   &motor control                                     &910    \\        
			susan          &MiBench    &image process algorithm                           &2,140   \\        
			layer3         &MiBench    &mp3 audio lib                                     &2,233   \\        
			jcmarker       &MiBench    &jpeg compose algorithm                            &1,444   \\        
			jdmarker       &MiBench    &jpeg decompose algorithm                          &2,068   \\        
			jcphuff        &MiBench    &jpeg Huffman entropy encoding routines            &694    \\        
			gtk            &MiBench    &GTK plotting routines                             &949    \\        
			g72            &mediaBench &routines for G.721 and G.723 conversions          &608    \\        
			vga            &mediaBench &Driver for Borland Graphics Interface             &386    \\        
			stc            &mediaBench &pson Stylus-Color Printer-Driver                  &492    \\        
			\hline
		\end{tabular}
	}
\end{table}

\begin{table}[t!]
	\caption{Side channel detection: benchmark statistics.}
	\label{tbl:bench2}
	\centering
	\scalebox{.7}{
		\begin{tabular}{|llp{.8\linewidth}r|}
			\hline
			Name  &Source &Description &Loc   \\ 
			\hline\hline
			hash           &hpn-ssh    &hash function                                     &320    \\        
			encoder        &LibTomCrypt&hex encode a string                               &134    \\        
			chacha20       &LibTomCrypt&chacha20poly1305 cipher                           &776    \\        
			ocb            &LibTomCrypt&OCB implementation                                &377    \\        
			aes            &LibTomCrypt&AES implementation                                &1,838   \\        
			str2key        &openssl    &key prepare for des                               &385    \\        
			des	           &openssl    &des cipher                                        &1,051   \\        
			seed           &linux-tegra&seed cipher                                       &487    \\        
			camellia       &linux-tegra&camellia cipher                                   &1,324   \\        
			salsa          &linux-tegra&Salsa20 stream cipher                             &279    \\        
			\hline
		\end{tabular}
	}
\end{table}
	
\subsection{Effectiveness: Execution Time Estimation}

We first compare our method with the state-of-the-art, non-speculative
method~\cite{WuGSW18}.
%All the benchmarks we chose have been roughly inspected to contain
%conditional branches and different variables will be used in two
%branch bodies.
The results are shown in Table~\ref{tbl:wcet}.  For our method, we
also report the number of speculative cache misses (\#SpMiss), which
are not observable from outside of the CPU, the number of conditional
branches that can be speculatively executed, and the total number of
iterations of our method on loops.

The results show that our method detected more cache misses, thus
highlighting the unsoundness of the existing method and the
importance of modeling speculative execution during execution time
estimation. 

As for the analysis time, our method completed all the benchmarks,
although it took a longer time than the non-speculative analysis due to
its focus on being always sound.
The reason why it took significantly longer for the \textit{gtk}
benchmark, in particular, is because the program has a large data size
(of nearly 3 MB), which led to a large number of variables to be
tracked in the abstract cache state.
%
% %% why didn't you do this already?   That will make the result table look a lot stronger!!!
%
%It should be optimized in the future by partition the cache model so we don't need
%to track variables unrelated to current function.
	
\begin{table}[t!]
		\caption{Execution time estimation: comparisons in terms of the analysis time and the number of cache misses. }
		\label{tbl:wcet}
		\centering
		\scalebox{.7}{
			\begin{tabular}{|l|rr|rrrrr|}
				\hline
				\multirow{2}{*}{Name}    &\multicolumn{2}{c|}{Non-speculative}  &\multicolumn{5}{c|}{Speculative} \\
				\cline{2-8}
				               &Time (s) &\#Miss  &Time (s)  &\#Miss &\#SpMiss &\#Branch  &\#Iteration\\ 
				\hline\hline
				adpcm          &0.98     &24      &12.70     &32     &17  &75  &173    \\  %motor control
				susan          &19.40    &17      &248.40    &26     &17  &113 &464    \\  %image process algorithm
				layer3         &7.24     &78      &65.54     &88     &35  &241 &374    \\  %mp3 audio lib
				jcmarker       &0.20     &22      &3.40      &26     &11  &37  &72     \\  %jpeg compose algorithm 
				jdmarker       &2.89     &21      &15.18     &78     &55  &193 &726   \\  %jpeg decompose algorithm 
				jcphuff        &0.03     &12      &0.44      &12     &13  &25  &32    \\  %jpeg Huffman entropy encoding routines
				gtk            &19.90    &16      &274.76    &19     &13  &77  &190    \\  %jpeg Huffman entropy encoding 
				g72            &0.16     &6       &0.94      &9      &4   &41  &79   \\%Common routines for G.721 and G.723 conversions
				vga            &0.05     &4       &0.06      &4      &3   &3   &3   \\  %Driver for Borland Graphics Interface
				stc            &0.13     &10      &0.96      &23     &14  &39  &105   \\ %pson Stylus-Color Printer-Driver
				\hline
			\end{tabular}
		}
	\end{table}

Table~\ref{tbl:wcet_merge} compares two merging strategies in terms of
the analysis time, the number of cache misses, the number of
speculative cache misses, and the number of iterations.  The result is
somewhat surprising in that although \emph{merging at rollback point}
is more aggressive than \emph{just-in-time merging}, the later is
actually faster.  The reason is because
merging the speculative state with the normal state right after the
rollback point may force the normal state to become a coarser-grained
over-approximation.  This can lead to a slower convergence to a
coarser fixed point, as shown by the data in Columns 5 and 9.
\shepherd{However, there are exceptions, indicating that 
\emph{optimal merging} in general is \emph{problem-specific}, and the 
accuracy depends on the combined effects of branches and loops
in a program.}

\begin{table}[t!]
\caption{Execution time estimation: comparisons of two strategies for merging speculative executions.}
\label{tbl:wcet_merge}
\centering
\scalebox{.7}{
\begin{tabular}{|l|rrrr|rrrr|}
	\hline
	\multirow{2}{*}{Name}     &\multicolumn{4}{c|}{Merging at rollback point}  &\multicolumn{4}{c|}{Just-in-time merging} \\
	\cline{2-9}
	&Time(s)    &\#Miss &\#SpMiss   &\#Ite &Time(s)    &\#Miss &\#SpMiss   &\#Ite\\ 
	\hline\hline
	adpcm          &14.40    &31  &25  &261  &12.70  &32  &17  &173  \\  %motor control
	susan          &405.70   &30  &29  &620  &248.40 &26  &17  &464  \\  %image process algorithm
	layer3         &84.64    &94  &53  &471  &65.54  &88  &35  &374  \\  %mp3 audio lib
	jcmarker       &4.80     &27  &19  &99   &3.40   &26  &11  &72   \\  %jpeg compose algorithm 
	jdmarker       &16.11    &35  &59  &777  &15.18  &78  &55  &726  \\  %jpeg decompose algorithm 
	jcphuff        &0.48     &12  &10  &36   &0.44   &12  &13  &32   \\  %jpeg Huffman entropy encoding routines
	gtk            &358.56   &24  &26  &236  &274.76 &19  &13  &190  \\  %jpeg Huffman entropy encoding 
	g72            &1.28     &7   &1   &122  &0.94   &9   &4   &79   \\  %G.721 and G.723 conversions
	vga            &0.07     &4   &3   &5    &0.06   &4   &3   &3      \\ 
	stc            &1.86     &31  &35  &222  &0.96   &23  &14  &105   \\ 
	\hline
\end{tabular}
}
\end{table}

\subsection{Effectiveness: Side Channel Detection}

Table~\ref{tbl:sidechannel} shows the results for side channel
detection, including comparisons of the two methods in terms of the
analysis time and whether leaks are detected.
In this context, a leak refers to the dependency between the cache
behavioral difference and sensitive data; furthermore, whether there
is a leak or not often depends on the input buffer size controlled by
the (potentially malicious) user.  Thus, during experiments, we set
the buffer size to various values from 32K bytes (the size of cache we
use) down to 0 byte.

Generally speaking, the larger the buffer size, the easier
that the client program triggers the behavioral difference.
Thus we first set the buffer size to 32KB, and starting from there we 
gradually reduce the buffer size and keep track of the impact of speculative execution
on cache state, until the two methods return different results.

Since the benchmarks are mostly cryptographic algorithms, which are
relatively small in terms of the number of lines of code, the analysis time is
short.  Furthermore, our method successfully detected leaks in half of
the benchmarks, whereas the existing (unsound) method did not detect
leaks in any of them. 
This highlights the importance of having a sound static cache analysis
for speculative execution, e.g., to detect more leaks and avoid
bogus proofs (that there is no leak).
On one of the benchmarks, \emph{des}, leaks are detected even if the
buffer size is set to 0 because, even without the client
program, the benchmark program itself has a user controlled buffer,
which can be set to sizes that induce timing side-channel leaks under
speculative execution.

\shepherd{
As a static analysis procedure, our method may generate false
positives.  In addition to abstraction, the other source of false
positives is modeling of the speculative execution.  Therefore, for each of the
new leaks detected by our method in Table~\ref{tbl:sidechannel}, we
manually inspected the software code and the execution trace.
Our inspection confirmed that all of them are indeed real leaks; that is,
there exist specific memory/cache layouts and execution traces that
induce the leaks.}

\begin{table}[t!]
\caption{Side channel detection: comparisons in terms of the analysis time and whether leaks are detected.}
\label{tbl:sidechannel}
\centering
\scalebox{0.7}{
\begin{tabular}{|l|c|rc|rc|}
	\hline
	\multirow{2}{*}{Name} & \multirow{2}{*}{Buffer (byte)}    &\multicolumn{2}{c|}{Non-speculative}  &\multicolumn{2}{c|}{Speculative} \\\cline{3-6}
	            &           &Time (s) &Leak Detected   &Time (s) &Leak Detected \\ 
\hline\hline
 hash           &31,808     &0.67    &No               &1.15     &Yes   \\  %hash function
\hline             
encoder   	   &32,512      &0.03    &No               &0.10	 &Yes   \\  %hex encode a string
\hline             
chacha20       &26,304      &1.18    &No               &9.24     &Yes   \\  %chacha20poly1305 cipher	
\hline             
ocb            &31,616      &0.10    &No               &0.68     &Yes   \\  %OCB implementation
\hline             
aes            &32,768      &0.08    &No               &2.13     &No   \\  %hex encode a string	
\hline             
str2key        &32,768      &0.01    &No               &0.01     &No   \\  %key prepair for des
\hline             
des			   &0           &0.60    &No               &14.20    &Yes   \\  %des cipher
\hline         
seed           &32,768      &0.01    &No               &0.07     &No   \\  %seed cipher
\hline         
camellia       &32,768      &0.35    &No               &6.35     &No   \\  %camellia cipher
\hline         
salsa          &32,768      &0.02    &No               &0.06     &No   \\  %camellia cipher
\hline
\end{tabular}
}
\end{table}

%\subsection{Threat to Validity}
%\label{sec:discussion}
%1. 

%\cwnote{Rebuttal to the (anticipated) questions of the reviewers..}	

%\newpage	
\section{Related Work}
\label{sec:related}

Abstract interpretation~\cite{cousot1977abstract} is a framework for
conducting static analysis and proving properties.  Ferdinand and
Wilhelm~\cite{ferdinand1999cache,ferdinand1999} pioneered the use of
abstract interpretation in may- and must-hit cache
analyses~\cite{wilhelm2010static}. Others also used similar techniques to
detect timing side
channels~\cite{DoychevFKMR13,BartheKMO14,WuGSW18}. However, prior
works focused primarily on improving abstract interpretation without
considering speculative execution.

There are some techniques that consider the impact of speculative
execution~\cite{li2006modeling}, but only for the instruction
pipeline. In a commercial tool named \emph{AIT}, speculations are also
considered during execution time estimation by leveraging a
standalone pipeline analysis as a driver~\cite{wilhelm2010static}.
Since the tool is propriety, details of this analysis have not been
made public; therefore, it is not clear how 
speculative execution is modeled during abstract interpretation.

Our method differs from the large body of work on statistically
estimating the worst-case execution time of real-time
software~\cite{MitraTT18a,LiMR03,LiSLMR09} using either CPU simulators
or characteristics of prior simulation results~\cite{theiling2000}.
These techniques, while useful, are not designed to be sound,
and hence may not be suitable for the applications
that we have in mind, such as detecting side-channel leaks or proving
that leaks do not exist. The reason is because, if the analysis is not
sound, the proof may not be valid and as a consequence, leaks may be
left undetected.

For timing side channels, many analysis and verification
techniques~\cite{PasareanuPM16,BultanYAA17,BangAPPB16,BrennanSB18,PhanBPMB17,WuGSW18,SungPW18,GuoWW18}
have been developed, including the one proposed by Chen et
al.~\cite{ChenFD17}, which uses Cartesian Hoare Logic~\cite{SousaD16}
to prove that timing leaks of a program are bounded.  Antonopoulos
et al.~\cite{AntonopoulosGHK17} also developed a method for proving
the absence of timing channels.  However, these methods only
consider \emph{instruction-induced} timing variance while ignoring the
cache.

There are also techniques for improving the accuracy of cache
analysis, e.g., by using symbolic execution or model checking to
refine the cache analysis
results~\cite{chuprecise,metta2016tic,touzeau2017} and by extending
the analysis from single-core to multi-core
CPUs~\cite{lv2010combining}.  However, none of these techniques
considered speculative execution, which is the focus of our work.

%In this paper, we have explored the use of static cache analysis for
%timing side channel detection.  Although there is a large body of work
%on detecting timing side channels using static
%analysis~\cite{CacheAuditpaper,UTAustinpapers,JeffFrosterPapers}, none
%of them considered speculative execution.  We show that, under
%speculative execution, these existing method may be unsound.

\section{Conclusions}
\label{sec:conclusion}

We have presented a new abstract interpretation technique that can
soundly analyze a program under speculative execution.  The goal is to
lift existing static analyzers, which were geared toward analyzing only
non-speculative executions, so that they become sound also for
speculative executions.  We have implemented the technique in a static
cache analysis tool and evaluated it on two sets of benchmarks, for
execution time estimation and side channel detection.  Our
experimental results show that the method can detect many cache misses
and side-channel leaks overlooked by a state-of-the-art non-speculative
analysis technique.

\section*{Acknowledgments}

This work was partially funded by the U.S.\ National Science
Foundation (NSF) under the grants CNS-1617203 and CNS-1702824 and the
Office of Naval Research (ONR) under the grant N00014-17-1-2896.
%
%Any opinions, findings, and conclusions expressed in this material are
%those of the authors and do not necessarily reflect the views of the
%funding agencies.

\newpage\clearpage
\bibliography{spectre}

\newpage\clearpage
\appendix
\appendixpage

We review the original cache analysis in Appendix~\ref{app:original}
and then present the optimized analysis in
Appendix~\ref{app:modified}. Finally, we compare them in
Appendix~\ref{app:comparison}.

\section{The Original Cache Analysis}
\label{app:original}

\subsection{The Cache State}

Recall that, without the shadow variables, the abstract cache state is
defined as $S = \langle Age(v_1), \ldots, Age(v_n) \rangle$, where
each regular variable $v_i$ ($1\leq i\leq n$) has an age $Age(v_i)$.
For our \emph{must-hit} analysis, $Age(v_i)$ is an upper bound of the
oldest cache line age for $v_i$ along all program paths.

\subsubsection{The Transfer Function}
\label{app:transfer_original}

Given a state $S$ and an instruction $inst$, $\proc{Transfer}(S,inst)$
computes the new state $S' = \langle
\mathit{Age'}(v_1),\ldots,\mathit{Age'}(v_n) \rangle$ as follows:
\begin{itemize}
\item If $inst$ does not access memory at all, then $S' = S$.
\item Otherwise, let $v\in V$ be the variable accessed; $S'$ is
  computed as follows:
  \begin{itemize}
  \item 
    For the accessed variable $v$, set $\mathit{Age}'(v) = 1$ in $S'$.
  \item 
    For variable $u \in V$ whose age may be 
    younger than $v$ in $S$, increment the age of $u$;
    that is, \\ 
    $\mathit{Age}(u)$ \textcolor{red}{$<$}
    $\mathit{Age}(v) \rightarrow \mathit{Age'}(u) = \mathit{Age}(u)
    +1$.
  \item 
    For any other variable $w \in V$, set $\mathit{Age'}(w) =
    \mathit{Age}(w)$.
  \end{itemize}
\end{itemize}

%\textcolor{darkgreen}
{
\begin{example}
On the left-hand side of Figure~\ref{fig:cacheupdate}, $v$ is the
variable accessed, while $u_1$--$u_4$ are younger than $v$ originally.
Thus, executing $inst$ increases the ages of $u_1$--$u_4$ by 1.\\
On the right-hand side, $u$ is younger than $v$ originally, while
$w_1$ and $w_2$ are older than $v$.  Thus, executing $inst$ increases
the age of $u$ by 1, while the ages of $w_1$ and $w_2$ remain
unchanged.
\end{example}
}

\subsubsection{The Join Operation}

A variable must be in the cache \emph{only if} it is in the cache
according to both cache states before the join.  Therefore, given two
cache states $S = \langle Age(v_1),\ldots,Age(v_n) \rangle$ and $S' =
\langle Age(v'_1),\dots,Age(v'_n) \rangle$, we define $S'' = S \sqcup
S'$ as follows: $S'' = \langle max(Age(v_1),Age(v'_1)), \ldots,
max(Age(v_n),Age(v'_n)) \rangle$.

%\textcolor{darkblue}
{
\begin{example}
For the example in Figure~\ref{fig:cachejoin}, the merged state does
not have $y$ (or $t$) because, in at least one of the two states, $y$
(or $t$) was already outside of the cache. Furthermore, $x,z$ and $k$
appear in their oldest possible cache lines.
\end{example}
}

\section{Cache Analysis with Shadow Variables}
\label{app:modified}

\subsection{The Cache State}

To improve the accuracy, for each program variable $v_i$ ($1\leq i\leq
n$), we add a shadow variable $\exists v_i$ to represent the youngest
cache line that holds $v_i$ along some program path.
With the shadow variables, the new cache state is defined as $S =
\langle Age(v_1), \ldots, Age(v_n)$, $Age(\exists v_1),\dots$,
$Age(\exists v_n) \rangle$.

%\textcolor{darkblue}
{
\begin{example}
For the example in Figure~\ref{fig:cachejoin}, the original state is
$[\{~\},\{~\},\{x,z\},k]$, but the modified state with shadow
variables is $[\{\exists x, \exists t\}, \{\exists y, \exists z\},
  \{x,z\}, \{\exists k, k\}]$.  Here, $\exists x$ is in the youngest
cache line for $x$ along some path whereas $x$ is in the oldest
possible cache line along all paths.\\
%
%The modified state contains both $\exists k$ and $k$, which happen to
%be in the same cache line --- while it may be tempting, we should not
%remove the shadow variable $\exists k$; otherwise, the algorithm would
%not work properly. \\\\
%
The modified state can be viewed as two separate states: $S^{MUST}
= [ \{\}, \{\}, \{x,z\}, \{k\} ]$, the oldest cache lines along all
paths, and $S^{MAY} = [ \{\exists x, \exists t\}, \{\exists y, \exists
  z\}, \{\}, \{\exists k\} ]$, the youngest cache lines along some
paths.
\end{example}
}

\subsubsection{The Transfer Function}

Given the state $S$ and an instruction $inst$, the new state $S'$ is
computed using the modified $\proc{Transfer}(S,inst)$ as follows:

\begin{itemize}
%\item 
%  If no shadow variable is in the cache, i.e., $\mathit{Age}(\exists
%  v) = N+1$ in $S$ for all variable of the form $\exists v$, the ages
%  of regular variables are updated using the original transfer
%  function in Appendix~\ref{appendix:transfer_original}.
\item
  If $inst$ does not access memory at all, then $S' = S$. 

\item Otherwise, let $v\in V$ be the variable accessed in $S$;  $S'$ is computed
  as follows:

  \begin{itemize}
  \item 
  For the accessed variable $v$, set $Age'(v)=1$ in $S'$; 
  \textcolor{purple}{
  and then update the shadow variables (in $S^{MAY}$):} \textcolor{blue}{
  \begin{itemize}
    \item
    Set $Age'(\exists v) = 1$.
    \item
    If $Age(\exists u)$ \textcolor{red}{$\leq$} $Age(\exists v)$, set $Age'(\exists u) =
    Age(\exists u)+1$.
    \item
    If $Age(\exists u) > Age(\exists v)$, set $Age'(\exists u) = Age(\exists u)$.
  \end{itemize}
  }

  \item
    For variable $u\in V$ whose age may be younger than $v$ in $S$, increment  the age of $u$ 
    \textcolor{purple}{
    --- unless the number of variables younger than or equal to $u$, denoted $N_{Young}(u)$, is smaller than $Age(u)$.} \textcolor{blue}{
      \begin{enumerate}
      \item 
        $N_{Young}(u) =|\{ \exists w |~ Age'(\exists w) \leq Age(u) \wedge w\neq u \}|$.
      \item 
        If $N_{Young}(u) \geq  Age(u)$, then  $Age'(u)=Age(u)+1$;
      \item 
        else, $Age'(u)=Age(u)$;
      \end{enumerate}
    }
  \item 
    For any other variable $w \in V$,  set $Age'(w) =
    Age(w)$.
  \end{itemize}
\end{itemize}

There are two steps: we first compute the $S^{MAY}$ set of shadow
variables and then compute the $S^{MUST}$ set of regular variables.
$N_{young}(u)$ is computed based on $S^{MAY}$ and used to refine the
aging rule for $u$ when computing $S^{MUST}$.

The condition for aging $\exists u$, i.e., $Age(\exists
u)$\textcolor{red}{$\leq$}$Age(\exists v)$, also differs from the one
for aging $u$, which is $Age(u)$\textcolor{red}{$<$}$Age(v)$.

\begin{example}
%\textcolor{darkblue}
{
Below is the comparison of the original and modified transfer
functions for the example in Figure~\ref{fig:cachejoin}.
}

\vspace{1ex}
\noindent
\scalebox{0.7}{
\begin{tabular}{|l|l|c|l|} \hline
          & State $S$                   & Instruction &  New State $S'$      \\\hline

original  & $[\{\},\{\},\{x,z\},k]$ &  ref x      &  $[x,\{\},z,k]$ \\
          &                             &             &                              \\\hline
original  & $[\{\},\{\},\{x,z\},k]$ &  ref z      &  $[z,\{\},x,k]$ \\
          &                             &             &                              \\\hline
original  & $[\{\},\{\},\{x,z\},k]$ &  ref k      &  $[k,\{\},\{\},\{x,z\}]$ \\
          &                             &             &                              \\\hline
original  & $[\{\},\{\},\{x,z\},k]$ &  ref y      &  $[y,\{\},\{\},\{x,z\}]$ \\
          &                             &             &                              \\\hline
original  & $[\{\},\{\},\{x,z\},k]$ &  ref t      &  $[t,\{\},\{\},\{x,z\}]$ \\
          &                             &             &                              \\\hline
\hline
modified  & \textcolor{purple}{$[\{\exists x,\exists t\},\{\exists y,\exists z\},\{\},\{\exists k\}]$}  &  ref x      &  \textcolor{purple}{$[\{\exists x\},\{\exists t,\exists y,\exists z\},\{\},\{\exists k\}]$} \\
          & $[\{\},\{\},\{x,z\},k]$ &             &  $[\{x\},\{\},\{z\},\{k\}]$ \\\hline
modified  & \textcolor{purple}{$[\{\exists x,\exists t\},\{\exists y,\exists z\},\{\},\{\exists k\}]$}  &  ref z      &  \textcolor{purple}{$[\{\exists z\},\{\exists t,\exists y,\exists x\},\{\},\{\exists k\}]$} \\
          & $[\{\},\{\},\{x,z\},k]$ &             &  $[\{z\},\{\},\{x\},\{k\}]$ \\\hline
modified  & \textcolor{purple}{$[\{\exists x,\exists t\},\{\exists y,\exists z\},\{\},\{\exists k\}]$}  &  ref k      &  \textcolor{purple}{$[\{\exists k\},\{\exists x,\exists t\},\{\exists y,\exists z\},\{\}]$} \\
          & $[\{\},\{\},\{x,z\},k]$ &             &  $[\{k\},\{\},\{\},\{x,z\}]$ \\\hline
modified  & \textcolor{purple}{$[\{\exists x,\exists t\},\{\exists y,\exists z\},\{\},\{\exists k\}]$}  &  ref y      &  \textcolor{purple}{$[\{\exists y\},\{\exists x,\exists t\},\{\exists z\},\{\exists k\}]$} \\
          & $[\{\},\{\},\{x,z\},k]$ &             &  $[\{y\},\{\},\{\},\{x,z\}]$ \\\hline
modified  & \textcolor{purple}{$[\{\exists x,\exists t\},\{\exists y,\exists z\},\{\},\{\exists k\}]$}  &  ref t      &  \textcolor{purple}{$[\{\exists t\},\{\exists x,\exists y,\exists z\},\{\},\{\exists k\}]$} \\
          & $[\{\},\{\},\{x,z\},k]$ &             &  $[\{t\},\{\},\{\},\{x,z\}]$ \\\hline

\end{tabular}
}
\vspace{1ex}

\end{example}

\subsubsection{The Join Operation}

A regular variable $v\in V$ must be in the cache \emph{only if} it is
in the cache according to both states. Thus, after the join, its
age is the maximum of the two prior ages. 

A shadow variable $\exists v$ may be in the cache \emph if it is in
the cache according to either of the two states.  Thus, after the
join, its age is the minimum of the two prior ages. 

Formally, the joined state $S'' = S \sqcup S'$ is defined as follows:
$S'' = \langle max(Age(v_1),Age(v'_1))$, $\ldots$,
$max(Age(v_n),Age(v'_n))$, $\ldots$ $min(Age(\exists v_1),Age(\exists
v'_1))$, $\ldots$, $min(Age(\exists v_n),Age(\exists v'_n)) \rangle$.

%\textcolor{darkblue}
{
\begin{example}
For the example in Figure~\ref{fig:cachejoin}, the two states prior to
the join are 
$S  = [\{\exists x, x\}, \{\exists y, y\}, \{\exists z,  z\}, \{\exists k, k\}]$ and 
$S' = [\{\exists t, t\}, \{\exists z, z\}, \{\exists x, x\}, \{\exists k, k\}]$.  \\
The joined state is 
$S'' = [\{\exists x, \exists t\}, \{\exists y, \exists z\}, \{x,z\}, \{\exists k, k\}]$.
Again, this can be viewed as two separate sets: the must set $[ \{\},
  \{\}, \{x,z\}, \{k\} ]$ and may set $[ \{\exists x, \exists t\},
  \{\exists y, \exists z\}, \{\}, \{\exists k\} ]$.
\end{example}
}

\section{Comparison}
\label{app:comparison}

We apply both the original analysis and the modified analysis to the
example in Figure~\ref{fig:widening} after loop unrolling, and compare
the results.  In the table below, Column~2 shows the result of the
original analysis and Column~3 shows the result of the modified
analysis.

For ease of comprehension, here, we split the modified cache state $S$
into two parts: $S^{MAY}$, which has the shadow variables, and
$S^{MUST}$, which has the regular variables.
Figure~\ref{fig:existhit} shows the same states as in this table, but
in the combined form.

\vspace{1ex}
\noindent
\scalebox{0.7}{
\begin{tabular}{|l|r|r|} \hline
 State   & Cache state (original)           & Cache state (w/ shadow variables)   \\\hline

$S_0 =$ 
                 & $[ \bot  ,\bot, \bot, \bot ]$  & \textcolor{purple}{$[ \bot, \bot, \bot, \bot ]$}      \\
                 &                                & $[ \bot, \bot, \bot, \bot ]$      \\\hline
$S_1 =\proc{Trans}(S_0,$ ref a)     
                 & $[ a     ,\bot, \bot, \bot ]$  & \textcolor{purple}{$[ \{\exists a\}   , \bot, \bot, \bot ]$}      \\
                 &                                & $[ a   , \bot, \bot, \bot ]$      \\\hline
$S_2 =\proc{Trans}(S_1,$ ref b)
                 & $[ b     , a, \bot,\bot ]$     & \textcolor{purple}{$[ \{\exists b\}   , \{\exists a\}   , \bot, \bot ]$}         \\
                 &                                & $[ b   , a   , \bot, \bot ]$         \\\hline
$S_3 =\proc{Trans}(S_1,$ ref c)
                 & $[ c     , a, \bot,\bot ]$     & \textcolor{purple}{$[ \{\exists c\}   , \{\exists a\}   , \bot, \bot ]$}      \\
                 &                                & $[ c   , a   , \bot, \bot ]$      \\\hline
$S_4 =\proc{Join}(S_2,S_3$) 
                 & $[ \{\}, a, \bot,\bot ]$      & \textcolor{purple}{$[ \{\exists b, \exists c\}, \{\exists a\}   , \bot, \bot ]$} \\
                 &                               & $[ \{\}, a   , \bot, \bot ]$ \\\hline
$S_5 =\proc{Trans}(S_4,$  ref b)
                 & $[ b, \{\}, a, \bot ]$        & \textcolor{purple}{$[ \{\exists b\}, \{\exists a, \exists c\}, \bot,\bot ]$} \\
                 &                               & $[ b, \{\}, a,\bot ]$ \\\hline
$S_6 =\proc{Trans}(S_4,$  ref c)
                 & $[ c, \{\}, a, \bot ]$        & \textcolor{purple}{$[ \{\exists c\}, \{\exists a, \exists b\}, \bot,\bot ]$} \\
                 &                               & $[ c, \{\}, a,\bot ]$ \\\hline
$S_7 =\proc{Join}(S_5,S_6$) 
                 & $[ \{\}, \{\},a,\bot ]$       & \textcolor{purple}{$[ \{\exists b, \exists c\}, \{\exists a\}, \bot,\bot ]$} \\
                 &                               & $[ \{\}, \{\}, a,\bot ]$ \\\hline
$S_{8} =\proc{Trans}(S_7,$  ref b)
                 & $[ b, \{\}, \{\}, a ]$        & \textcolor{purple}{$[ \{\exists b\}, \{\exists a, \exists c\}, \{\},\bot ]$} \\
                 &                               & $[ b, \{\}, a,\bot ]$ \\\hline
$S_{9} =\proc{Trans}(S_7,$  ref c)
                 & $[ c, \{\}, \{\}, a ]$        & \textcolor{purple}{$[ \{\exists c\}, \{\exists a, \exists b\}, \{\},\bot ]$} \\
                 &                               & $[ c, \{\}, a,\bot ]$ \\\hline
$S_{10} =\proc{Join}(S_{8},S_{9}$) 
                 & $[ \{\}, \{\}, \{\},a ]$      & \textcolor{purple}{$[ \{\exists b, \exists c\}, \{\exists a\}, \{\},\bot ]$} \\
                 &                               & $[ \{\}, \{\}, a,\bot ]$ \\\hline
\end{tabular}
}
\vspace{1ex}

Initially, both $S^{MAY}$ and $S^{MUST}$ are
$[\bot,\bot,\bot,\bot]$.

Executing \texttt{ref a} will load $a$ into the first cache line.
Therefore, $S^{MAY} = [\{\exists a\},\bot,\bot,\bot]$ and $S^{MUST} =
[a,\bot,\bot,\bot]$.  The second part remains the same as the result
of the original analysis for this particular case, but in general, it
can be more accurate.

This is because, at each step, we first compute $S^{MAY}$ and then use
$N_{young}(u)$ computed from $S^{MAY}$ to more accurately compute
$S^{MUST}$.

\subsection{Same Results}

Sometimes, $S^{MUST}$ remains the same as the result of the original
analysis.

For instance, the original result for $S_5=\proc{Trans}(S_4,$ref b$)$
is $[b,\{\},a,\bot]$. In the new analysis, we first compute $S^{MAY} =
[\{\exists b\},\{\exists a,\exists c\},\bot,\bot]$ and then compute
$S^{MUST}$ as follows:
\begin{itemize}
\item $Age(a) = 2$ and $Age(b) = 5$ (outside of the cache).
\item $Age'(\exists b) = 1$, $Age'(\exists a) = 2$, and
  $Age'(\exists c) = 2$.
\item Thus, $N_{young}(a) = |\{\exists b, \exists c\}| = 2$.
\item Since $Age(a)<Age(b)$ and $N_{young}(a) \geq Age(a)$, we set
  $Age'(a) = Age(a)+1$.
\end{itemize}
Therefore, $S^{MUST} = [b,\{\},a,\bot]$.

\subsection{Better Results}

Sometimes, $S^{MUST}$ is more accurate than the result of the original
analysis.  

For instance, the original result for
$S_8=\proc{Trans}(s_7,$ref b$)$ is $[b,\{\},\{\},a]$.  In the new
analysis, we first compute $S^{MAY} = [\{\exists b\},\{\exists
  a,\exists c\},\bot,\bot]$ and then compute $S^{MUST}$ as follows:
\begin{itemize}
\item $Age(a) = 3$ and $Age(b) = 5$ (outside of the cache).
\item $Age'(\exists b) = 1$, $Age'(\exists a) = 2$, and
  $Age'(\exists c) = 2$.
\item Thus, $N_{young}(a) = |\{\exists b, \exists c\}| = 2$.
\item Since $Age(a)<Age(b)$ but $N_{young}(a) \not\geq Age(a)$, we set
  $Age'(a) = Age(a)$.
\end{itemize}
Therefore, $S^{MUST} = [b,\{\},a,\bot]$, which is more accurate than
$[b,\{\},\{\},a]$.

\subsection{Correctness}

In general, $S^{MUST}$ is either the same as, or more accurate than,
the result of the original analysis -- due to the following modified
rule for aging the variable $u$.

\textcolor{purple}{
Whenever $Age(u)<Age(v)$ but $N_{young}(u) < Age(u)$, we set $Age'(u)
= Age(u)$.
}

This optimization is safe because, if \emph{the number of shadow
  variables that are younger than or equal to $u$}, denoted
$N_{young}(u)$, is less than the current age of $u$, there must be
younger cache lines to hold all of them.  In such a case, we should
not (unnecessarily) increase the age of $u$.

\end{document}

%% file: fig/cacheupdate.tex
  \begin{tikzpicture}[fill=blue!20,font=\footnotesize]
  \draw (0, 0) rectangle (1.2, 0.4);
  \node[above=5pt, right] at (0.4, 0)   {$u_4$};
  \draw (0, 0.4) rectangle (1.2, 0.8);
  \node[above=5pt, right] at (0.4, 0.4) {$u_3$} ;	
  \draw (0, 0.8) rectangle (1.2, 1.2);
  \node[above=5pt, right] at (0.4, 0.8) {$u_2$};
  \draw (0, 1.2) rectangle (1.2, 1.6);
  \node[above=5pt, right] at (0.4, 1.2) {$u_1$};
  \draw (2, 0) rectangle (3.2, 0.4);
  \node[above=5pt, right] at (2.4, 0)   {$u_3$};
  \draw (2, 0.4) rectangle (3.2, 0.8);
  \node[above=5pt, right] at (2.4, 0.4) {$u_2$} ;	
  \draw (2, 0.8) rectangle (3.2, 1.2);
  \node[above=5pt, right] at (2.4, 0.8) {$u_1$};
  \draw (2, 1.2) rectangle (3.2, 1.6);
  \node[above=5pt, right] at (2.4, 1.2) {$v$};
  \draw[-stealth,black] (1.2, 1.4) -- (2, 1.0);
  \draw[-stealth,black] (1.2, 1.0) -- (2, 0.6);
  \draw[-stealth,black] (1.2, 0.6) -- (2, 0.2);
  \draw[-stealth,black] (3.8, 0.2) -- (3.8, 1.4) node [align =center,pos = 0.5,right] {young\\age};
   \draw (5, 0) rectangle (6.2, 0.4);
   \node[above=5pt, right] at (5.4, 0)   {$w_2$};
   \draw (5, 0.4) rectangle (6.2, 0.8);
   \node[above=5pt, right] at (5.4, 0.4) {$w_1$} ;	
   \draw (5, 0.8) rectangle (6.2, 1.2);
   \node[above=5pt, right] at (5.4, 0.8) {$v$};
   \draw (5, 1.2) rectangle (6.2, 1.6);
   \node[above=5pt, right] at (5.4, 1.2) {$u$};
   \draw (7, 0) rectangle (8.2, 0.4);
   \node[above=5pt, right] at (7.4, 0)   {$w_2$};
   \draw (7, 0.4) rectangle (8.2, 0.8);
   \node[above=5pt, right] at (7.4, 0.4) {$w_1$} ;	
   \draw (7, 0.8) rectangle (8.2, 1.2);
   \node[above=5pt, right] at (7.4, 0.8) {$u$};
   \draw (7, 1.2) rectangle (8.2, 1.6);
   \node[above=5pt, right] at (7.4, 1.2) {$v$};
   \draw[-stealth,black] (6.2, 1.4) -- (7, 1.0);
   \draw[-stealth,black] (6.2, 1.0) -- (7, 1.4);
   \draw[-stealth,black] (6.2, 0.6) -- (7, 0.6);
   \draw[-stealth,black] (6.2, 0.2) -- (7, 0.2);

  %  \node[above=5pt, right, color=blue] at (1, 2.8) {, tmp};

  \end{tikzpicture}
 

%% file: fig/cachejoin.tex
  \begin{tikzpicture}[fill=blue!20,font=\footnotesize]
  \draw (0, 0) rectangle (1.2, 0.3);
  \node[above=4pt, right] at (0.4, 0)   {k};
  \draw (0, 0.3) rectangle (1.2, 0.6);
  \node[above=4pt, right] at (0.4, 0.3) {z} ;	
  \draw (0, 0.6) rectangle (1.2, 0.9);
  \node[above=4pt, right] at (0.4, 0.6) {y};
  \draw (0, 0.9) rectangle (1.2, 1.2);
  \node[above=4pt, right] at (0.4, 0.9) {x};
  \draw (2, 0) rectangle (3.2, 0.3);
  \node[above=4pt, right] at (2.4, 0)   {k};
  \draw (2, 0.3) rectangle (3.2, 0.6);
  \node[above=4pt, right] at (2.4, 0.3) {x} ;	
  \draw (2, 0.6) rectangle (3.2, 0.9);
  \node[above=4pt, right] at (2.4, 0.6) {z};
  \draw (2, 0.9) rectangle (3.2, 1.2);
  \node[above=4pt, right] at (2.4, 0.9) {t};
  \draw (1.0, -1) rectangle (2.2, -0.7);
  \node[above=4pt, right] at (1.35, -1) { \{ \}};
  \draw (1, -1.3) rectangle (2.2, -1);
  \node[above=4pt, right] at (1.35, -1.3) {\{ \}} ;	
  \draw (1, -1.6) rectangle (2.2, -1.3);
  \node[above=4pt, right] at (1.2, -1.6) {\{x, z\}};
  \draw (1, -1.9) rectangle (2.2, -1.6);
  \node[above=4pt, right] at (1.4, -1.9) {k};
  \draw[-stealth,black] (0.6, 0) -- (1.4, -0.7);
  \draw[-stealth,black] (2.6, 0) -- (1.8, -0.7);
  \draw[-stealth,black] (3.4, 0.15) -- (3.4, 1.05) node [align =center,pos = 0.5,right] {young\\age};
  \end{tikzpicture}
 

%% file: fig/trace1.tex
  \begin{tikzpicture}[fill=blue!20,font=\footnotesize]
  \draw (2, 9) rectangle (3, 9.5);
  % \node[] at (2.4, 9.6) {\tiny br};
  % \node[] at (2.4, 9.4) {\tiny header};
  
  \draw (1, 6) rectangle (2, 8);
  % \node[above=2pt, right] at (0.49, 1.2) {\tiny spec bound};
  %
  \draw (3,6) rectangle (4, 8);
  % \node[above=2pt, right] at (3.99, 1.2) {\tiny spec bound};
  %
  \draw[-stealth,black] (2.4, 9) -- (1.6, 8);
  \draw[-stealth,black] (2.6, 9) --(3.3, 8);
  \draw[-stealth,black] (1.5, 6) -- (2.3, 5);
  \draw[-stealth,black] (3.5, 6) --(2.7, 5);
  \draw (2,4.5) rectangle (3, 5);
  % \node[] at (2.5, 4.8) {br};
  % \node[] at (2.5, 4.5) {tail};
  \draw[densely dotted]  (1, 7.75) -- (2, 7.75);
  \draw[densely dotted]  (1, 7.5) -- (2, 7.5);
  \draw[densely dotted]  (1, 7.25) -- (2, 7.25);
  \draw[densely dotted]  (1, 7) -- (2, 7);
  \draw[densely dotted]  (3, 7.75) -- (4, 7.75);
  \draw[densely dotted]  (3, 7.5) -- (4, 7.5);
  \draw[densely dotted]  (3, 7.25) -- (4, 7.25);
  \draw[densely dotted]  (3, 7) -- (4, 7);
  \node[] at (1.45, 7.85) {\tiny $inst_0$};
  \node[] at (1.45, 7.6) {\tiny $inst_1$};
  \node[] at (1.45, 7.35) {\tiny ...};
  \node[] at (1.45, 7.1) {\tiny $inst_B$};
  \node[] at (3.42, 7.85) {\tiny $inst_0$};
  \node[] at (3.42, 7.6) {\tiny $inst_1$};
  \node[] at (3.42, 7.35) {\tiny ...};
  \node[] at (3.42, 7.1) {\tiny $inst_B$};
  \draw [blue, -stealth] (2.1,9.6) to [out=-90,in=90] (2.1,9) to [out=-100,in=80] (1.1,8) to[out=-90,in=90] (1.1, 6) to [out=-80,in=100] (2.1, 5.1) to [out=-90,in=90] (2.1, 4.3);
   \draw [blue, -stealth] (2.9,9.6) to [out=-90,in=90] (2.9,9) to [out=-80,in=100] (3.9,8) to[out=-90,in=90] (3.9, 6) to [out=-100,in=80] (2.9, 5.1) to [out=-90,in=90] (2.9, 4.3);
   %
  %  \draw [red, dashed, ->] (2.7,10.2) to [out=-80,in=100] (2.8,8.8) to [out=-80,in=100] (3.9,8) to[out=-70,in=80] (3.8, 7.3) to [out=-115,in=-45] (2.2, 8.2) to [out=120,in=80] (1.9, 8) to[out=-100,in=100] (1.9, 6) to [out=-80,in=100] (2.4, 5.1) to [out=-90,in=90] (2.4, 3.7);
  %  %
  % \draw [red, dashed, ->] (2.6,10.2) to [out=-80,in=100] (2.7,8.8) to [out=-85,in=110] (3.8,8) to[out=-70,in=80] (3.7, 7.7) to [out=-115,in=-35] (2.3, 8.3) to [out=110,in=80] (1.8, 8) to[out=-100,in=100] (1.8, 6) to [out=-80,in=100] (2.3, 5.1) to [out=-90,in=90] (2.3, 3.7);
  % \draw [red, dashed, ->] (2.5,10.2) to [out=-80,in=100] (2.6,8.7) to [out=-85,in=110] (3.6,8) to[out=-70,in=80] (3.6, 7.9) to [out=-115,in=-35] (2.4, 8.4) to [out=110,in=80] (1.7, 8) to[out=-100,in=100] (1.7, 6) to [out=-80,in=100] (2.2, 5.1) to [out=-90,in=90] (2.2, 3.7);

   \draw [red, dashed, -stealth] (2.8,9.6) to [out=-90,in=90] (2.8,9)  to [out=-80,in=100] (3.8, 8) to [out=-90,in=70] (3.76, 7.5) to [out=-110,in=-70] (2.2, 8.0) to [out=110,in=60] (1.9, 8) to[out=-90,in=90] (1.9, 6) to [out=-70,in=90] (2.6, 5.1) to [out=-90,in=90] (2.6, 4.3);
  \draw [red, dashed, -stealth] (2.75,9.6) to [out=-90,in=90] (2.75,9)  to [out=-80,in=110] (3.75, 8) to [out=-90,in=80] (3.75, 7.7) to [out=-90,in=-60] (2.25, 8.15) to [out=120,in=60] (1.83, 8) to[out=-90,in=90] (1.83, 6) to [out=-80,in=100] (2.5, 5.1) to [out=-90,in=90] (2.5, 4.3);
  \draw [red, dashed, -stealth] (2.7,9.6) to [out=-90,in=90] (2.7,9)  to [out=-80,in=120] (3.7, 8) to [out=-90,in=90] (3.7, 7.95) to [out=-90,in=-60] (2.3, 8.3) to [out=160,in=80] (1.75, 8) to[out=-90,in=90] (1.75, 6) to [out=-90,in=110] (2.4, 5.1) to [out=-90,in=90] (2.4, 4.3);

\node[] at (2.15, 8.5) {\tiny T};
\node[] at (2.8, 8.5) {\tiny F};
  % \draw [dashed, ->] (4,1.2) to[out=180,in=-30] (2,2.5) to[out=-210,in=10] (1.1, 2.3);
  %
  % \draw (2, -0.9) rectangle (3.5, -0.5);
  % \node[above=5pt, right] at (2.2, -0.9) {br tail};
  % %
  % \draw (2.15, 1.5) rectangle (3.35, 2.1);
  % \node[above=2pt, right] at (2.15, 1.9) {\tiny if \$1==0;};
  % \node[above=2pt, right] at (2.35, 1.7) {\tiny    \$1=1;};
  % \node[above=2pt, right] at (2.15, 1.5) {\tiny else continue;};
  % %
  % \draw[->] (1, 0.5) -- (2.75, -0.5);
  % \draw[->] (4.5, 0.5) -- (2.75, -0.5);
  \end{tikzpicture}

%% file: fig/trace2.tex
  \begin{tikzpicture}[fill=blue!20,font=\footnotesize]
 \draw (2, 9) rectangle (3, 9.5);
  % \node[] at (2.5, 9.5) {br};
  % \node[] at (2.5, 9.2) {header};
  %
  \draw (1, 6) rectangle (2, 8);
  % \node[above=2pt, right] at (0.49, 1.2) {\tiny spec bound};
  %
  \draw (3,6) rectangle (4, 8);
  % \node[above=2pt, right] at (3.99, 1.2) {\tiny spec bound};
  %
  \draw[-stealth,black] (2.4, 9) -- (1.6, 8);
  \draw[-stealth,black] (2.6, 9) --(3.4, 8);
  \draw[-stealth,black] (1.5, 6) -- (2.3, 5);
  \draw[-stealth,black] (3.5, 6) --(2.7, 5);
  \draw (2,4.5) rectangle (3, 5);
  % \node[] at (2.5, 4.8) {br};
  % \node[] at (2.5, 4.5) {tail};
  \draw[densely dotted]  (1, 7.75) -- (2, 7.75);
  \draw[densely dotted]  (1, 7.5) -- (2, 7.5);
  \draw[densely dotted]  (1, 7.25) -- (2, 7.25);
  \draw[densely dotted]  (1, 7) -- (2, 7);
  \draw[densely dotted]  (3, 7.75) -- (4, 7.75);
  \draw[densely dotted]  (3, 7.5) -- (4, 7.5);
  \draw[densely dotted]  (3, 7.25) -- (4, 7.25);
  \draw[densely dotted]  (3, 7) -- (4, 7);
  \node[] at (1.45, 7.85) {\tiny $inst_0$};
  \node[] at (1.45, 7.6) {\tiny $inst_1$};
  \node[] at (1.45, 7.35) {\tiny ...};
  \node[] at (1.45, 7.1) {\tiny $inst_B$};
  \node[] at (3.42, 7.85) {\tiny $inst_0$};
  \node[] at (3.42, 7.6) {\tiny $inst_1$};
  \node[] at (3.42, 7.35) {\tiny ...};
  \node[] at (3.42, 7.1) {\tiny $inst_B$};
  \draw [blue, -stealth] (2.5,9.6) to [out=-90,in=90] (2.5,9) to [out=-100,in=90] (1.1,8) to[out=-90,in=90] (1.1, 6) to [out=-90,in=100] (2.5, 5.1) to [out=-90,in=90] (2.5, 4.3);
   \draw [blue, -stealth] (2.5,9.6) to [out=-90,in=90] (2.5,9) to [out=-80,in=90] (3.9,8) to[out=-90,in=90] (3.9, 6) to [out=-90,in=80] (2.5, 5.1) to [out=-90,in=90] (2.5, 4.3);

  \draw [red, dashed] (3.6,8.39) to [out=-20,in=90] (3.8, 7.5) to [out=-90,in=-70] (2.2, 8.0) to [out=110,in=60] (1.9, 8) to[out=-90,in=90] (1.9, 6) to [out=-70,in=90] (2.5, 5.1);
  \draw [red, dashed] (3.6,8.39) to [out=-30,in=90] (3.75, 7.7) to [out=-90,in=-60] (2.25, 8.15) to [out=120,in=60] (1.83, 8) to[out=-90,in=90] (1.83, 6) to [out=-80,in=100] (2.5, 5.1);
  \draw [red, dashed] (3.6,8.39) to [out=-40,in=90] (3.72, 7.95) to [out=-90,in=-60] (2.3, 8.3) to [out=160,in=80] (1.75, 8) to[out=-90,in=90] (1.75, 6) to [out=-90,in=110] (2.5, 5.1);

  \node[circle,fill=blue,inner sep=0pt,minimum size=3pt] at (2.5, 5.1) {};

\node[] at (2.15, 8.5) {\tiny T};
\node[] at (2.85, 8.5) {\tiny F};
  % \draw [dashed, ->] (4,1.2) to[out=180,in=-30] (2,2.5) to[out=-210,in=10] (1.1, 2.3);
  %
  % \draw (2, -0.9) rectangle (3.5, -0.5);
  % \node[above=5pt, right] at (2.2, -0.9) {br tail};
  % %
  % \draw (2.15, 1.5) rectangle (3.35, 2.1);
  % \node[above=2pt, right] at (2.15, 1.9) {\tiny if \$1==0;};
  % \node[above=2pt, right] at (2.35, 1.7) {\tiny    \$1=1;};
  % \node[above=2pt, right] at (2.15, 1.5) {\tiny else continue;};
  % %
  % \draw[->] (1, 0.5) -- (2.75, -0.5);
  % \draw[->] (4.5, 0.5) -- (2.75, -0.5);
  \end{tikzpicture}

%% file: fig/trace3.tex
  \begin{tikzpicture}[fill=blue!20,font=\footnotesize]
  \draw (2, 9) rectangle (3, 9.5);
  % \node[] at (2.4, 9.6) {\tiny br};
  % \node[] at (2.4, 9.4) {\tiny header};
  
  \draw (1, 6) rectangle (2, 8);
  % \node[above=2pt, right] at (0.49, 1.2) {\tiny spec bound};
  %
  \draw (3,6) rectangle (4, 8);
  % \node[above=2pt, right] at (3.99, 1.2) {\tiny spec bound};
  %
  \draw[-stealth,black] (2.4, 9) -- (1.6, 8);
  \draw[-stealth,black] (2.6, 9) --(3.4, 8);
  \draw[-stealth,black] (1.5, 6) -- (2.3, 5);
  \draw[-stealth,black] (3.5, 6) --(2.7, 5);
  \draw (2,4.5) rectangle (3, 5);
  % \node[] at (2.5, 4.8) {br};
  % \node[] at (2.5, 4.5) {tail};
  \draw[densely dotted]  (1, 7.75) -- (2, 7.75);
  \draw[densely dotted]  (1, 7.5) -- (2, 7.5);
  \draw[densely dotted]  (1, 7.25) -- (2, 7.25);
  \draw[densely dotted]  (1, 7) -- (2, 7);
  \draw[densely dotted]  (3, 7.75) -- (4, 7.75);
  \draw[densely dotted]  (3, 7.5) -- (4, 7.5);
  \draw[densely dotted]  (3, 7.25) -- (4, 7.25);
  \draw[densely dotted]  (3, 7) -- (4, 7);
  \node[] at (1.45, 7.85) {\tiny $inst_0$};
  \node[] at (1.45, 7.6) {\tiny $inst_1$};
  \node[] at (1.45, 7.35) {\tiny ...};
  \node[] at (1.45, 7.1) {\tiny $inst_B$};
  \node[] at (3.42, 7.85) {\tiny $inst_0$};
  \node[] at (3.42, 7.6) {\tiny $inst_1$};
  \node[] at (3.42, 7.35) {\tiny ...};
  \node[] at (3.42, 7.1) {\tiny $inst_B$};
  \draw [blue, -stealth] (2.5,9.6) to [out=-90,in=90] (2.5,9) to [out=-100,in=90] (1.1,8) to[out=-90,in=90] (1.1, 6) to [out=-90,in=100] (2.5, 5.1) to [out=-90,in=90] (2.5, 4.3);
   \draw [blue, -stealth] (2.5,9.6) to [out=-90,in=90] (2.5,9) to [out=-80,in=90] (3.9,8) to[out=-90,in=90] (3.9, 6) to [out=-90,in=80] (2.5, 5.1) to [out=-90,in=90] (2.5, 4.3);
   %

  %  \draw [red, dashed] (3.6,8.34) to [out=-30,in=90] (3.8,7.7) to[out=-70,in=70] (3.76, 7.3) to [out=-110,in=-60] (2.2, 8.2);
  % \draw [red, dashed] (3.6,8.34) to [out=-30,in=90] (3.7, 7.8) to [out=-110,in=-60] (2.2, 8.2);
  % \draw [red, dashed] (3.6,8.34) to [out=-30,in=90] (3.6,8) to[out=-110,in=-30] (2.2, 8.2) to [out=120,in=80] (1.7, 8) to[out=-100,in=100] (1.7, 6) to [out=-80,in=100] (2.5, 5.1);
 
  \draw [red, dashed] (3.6,8.39) to [out=-20,in=90] (3.8, 7.5) to [out=-90,in=-70] (2.2, 8.2);
  \draw [red, dashed] (3.6,8.39) to [out=-30,in=90] (3.75, 7.7) to [out=-90,in=-50] (2.2, 8.2)  to [out=150,in=60] (1.83, 8) to[out=-90,in=90] (1.83, 6) to [out=-70,in=90] (2.5, 5.1);
  \draw [red, dashed] (3.6,8.39) to [out=-40,in=90] (3.72, 7.95) to [out=-90,in=-30] (2.2, 8.2);

   \node[circle,fill=red,inner sep=0pt,minimum size=3pt] at (2.2, 8.2) {};
     \node[circle,fill=blue,inner sep=0pt,minimum size=3pt] at (2.5, 5.1) {};

\node[] at (2.15, 8.5) {\tiny T};
\node[] at (2.85, 8.5) {\tiny F};
  % \draw [dashed, ->] (4,1.2) to[out=180,in=-30] (2,2.5) to[out=-210,in=10] (1.1, 2.3);
  %
  % \draw (2, -0.9) rectangle (3.5, -0.5);
  % \node[above=5pt, right] at (2.2, -0.9) {br tail};
  % %
  % \draw (2.15, 1.5) rectangle (3.35, 2.1);
  % \node[above=2pt, right] at (2.15, 1.9) {\tiny if \$1==0;};
  % \node[above=2pt, right] at (2.35, 1.7) {\tiny    \$1=1;};
  % \node[above=2pt, right] at (2.15, 1.5) {\tiny else continue;};
  % %
  % \draw[->] (1, 0.5) -- (2.75, -0.5);
  % \draw[->] (4.5, 0.5) -- (2.75, -0.5);
  \end{tikzpicture}

%% file: fig/trace4.tex
  \begin{tikzpicture}[fill=blue!20,font=\footnotesize]
  \draw (2, 9) rectangle (3, 9.5);
  % \node[] at (2.4, 9.6) {\tiny br};
  % \node[] at (2.4, 9.4) {\tiny header};
  
  \draw (1, 6) rectangle (2, 8);
  % \node[above=2pt, right] at (0.49, 1.2) {\tiny spec bound};
  %
  \draw (3,6) rectangle (4, 8);
  % \node[above=2pt, right] at (3.99, 1.2) {\tiny spec bound};
  %
  \draw[-stealth,black] (2.4, 9) -- (1.6, 8);
  \draw[-stealth,black] (2.6, 9) --(3.4, 8);
  \draw[-stealth,black] (1.5, 6) -- (2.3, 5);
  \draw[-stealth,black] (3.5, 6) --(2.7, 5);
  \draw (2,4.5) rectangle (3, 5);
  % \node[] at (2.5, 4.8) {br};
  % \node[] at (2.5, 4.5) {tail};
  \draw[densely dotted]  (1, 7.75) -- (2, 7.75);
  \draw[densely dotted]  (1, 7.5) -- (2, 7.5);
  \draw[densely dotted]  (1, 7.25) -- (2, 7.25);
  \draw[densely dotted]  (1, 7) -- (2, 7);
  \draw[densely dotted]  (3, 7.75) -- (4, 7.75);
  \draw[densely dotted]  (3, 7.5) -- (4, 7.5);
  \draw[densely dotted]  (3, 7.25) -- (4, 7.25);
  \draw[densely dotted]  (3, 7) -- (4, 7);
  \node[] at (1.45, 7.85) {\tiny $inst_0$};
  \node[] at (1.45, 7.6) {\tiny $inst_1$};
  \node[] at (1.45, 7.35) {\tiny ...};
  \node[] at (1.45, 7.1) {\tiny $inst_B$};
  \node[] at (3.42, 7.85) {\tiny $inst_0$};
  \node[] at (3.42, 7.6) {\tiny $inst_1$};
  \node[] at (3.42, 7.35) {\tiny ...};
  \node[] at (3.42, 7.1) {\tiny $inst_B$};
  \draw [blue, -stealth] (2.5,9.6) to [out=-90,in=90] (2.5,9) to [out=-100,in=90] (1.1,8) to[out=-90,in=90] (1.1, 6) to [out=-90,in=100] (2.5, 5.1) to [out=-90,in=90] (2.5, 4.3);
   \draw [blue, -stealth] (2.5,9.6) to [out=-90,in=90] (2.5,9) to [out=-80,in=90] (3.9,8) to[out=-90,in=90] (3.9, 6) to [out=-90,in=80] (2.5, 5.1) to [out=-90,in=90] (2.5, 4.3);
  \draw [red, dashed] (3.6,8.39) to [out=-20,in=90] (3.8, 7.54) to [out=-90,in=-30] (1.6, 8.46);
  \draw [red, dashed] (3.6,8.39) to [out=-30,in=90] (3.75, 7.71) to [out=-90,in=-15] (1.6, 8.46);
  \draw [red, dashed] (3.6,8.39) to [out=-40,in=90] (3.72, 7.96) to [out=-90,in=0] (1.6, 8.46);

  \node[circle,fill=blue,inner sep=0pt,minimum size=3pt] at (1.6, 8.46) {};

\node[] at (2.15, 8.5) {\tiny T};
\node[] at (2.85, 8.5) {\tiny F};

  \end{tikzpicture}

%% file: fig/merge.tex
\scalebox{1.2}{
\begin{tikzpicture}[fill=blue!20,font=\footnotesize]
\draw (1.75, 3.5) rectangle (3, 3.9);
\node[above=5pt, right] at (1.75, 3.5) {\tiny 1: load a,b,c};
\draw (1.25, 2.5) rectangle (2.2, 2.9);
\node[above=5pt, right] at (1.25, 2.5) {\tiny 2: load d};
\draw (2.3, 2.5) rectangle (3.25, 2.9);
\node[above=5pt, right] at (2.3, 2.5) {\tiny 3: load e};
\draw[-stealth,black] (2.375, 3.5) -- (1.875, 2.9);
\draw[-stealth,black] (2.375, 3.5) -- (2.775, 2.9);
\node[left] at (2.2, 3.2) {\tiny T};
\node[right] at (2.5, 3.2) {\tiny F};
\draw (1.75, 1.5) rectangle (3, 1.9);
\node[above=5pt, right] at (1.87, 1.5) {\tiny4: load a};
\draw[-stealth,black] (1.875, 2.5) -- (2.37, 1.9);
\draw[-stealth,black] (2.775, 2.5) -- (2.38, 1.9);
\draw (5.7, 4) rectangle (6.5, 4.2);
\node[above=4pt, right] at (5.9, 3.95) {};
\draw (5.7, 4.2) rectangle (6.5, 4.4);
\node[above=4pt, right] at (5.9, 4.15) {\tiny a} ;	
\draw (5.7, 4.4) rectangle (6.5, 4.6);
\node[above=4pt, right] at (5.9, 4.35) {\tiny b};
\draw (5.7, 4.6) rectangle (6.5, 4.8);
\node[above=4pt, right] at (5.9, 4.55) {\tiny c};
\draw[-stealth,black, densely dotted]  (6.1, 4) -- (6.8, 3.3);
\draw[-stealth,black, densely dotted]  (6.1, 4) -- (7.8, 3.3);
\node[left] at (6.6, 3.6) {\tiny{T$_s$}};
\node[right] at (7.1, 3.6) {\tiny{F$_s$}};
\draw[-stealth,black]  (6.1, 4) -- (4.4, 3.3);
\draw[-stealth,black]  (6.1, 4) -- (5.4, 3.3);
\node[left] at (5, 3.6) {\tiny T};
\node[right] at (5.7, 3.6) {\tiny F};
\draw (4, 2.5) rectangle (4.8, 2.7);
\node[above=4pt, right] at (4.2, 2.45) {\tiny a};
\draw (4, 2.7) rectangle (4.8, 2.9);
\node[above=4pt, right] at (4.2, 2.65) {\tiny b} ;	
\draw (4, 2.9) rectangle (4.8, 3.1);
\node[above=4pt, right] at (4.2, 2.85) {\tiny c};
\draw (4, 3.1) rectangle (4.8, 3.3);
\node[above=4pt, right] at (4.2, 3.05) {\tiny d};
\draw (5, 2.5) rectangle (5.8, 2.7);
\node[above=4pt, right] at (5.2, 2.45){\tiny a};
\draw (5, 2.7) rectangle (5.8, 2.9);
\node[above=4pt, right] at (5.2, 2.65) {\tiny b} ;	
\draw (5, 2.9) rectangle (5.8, 3.1);
\node[above=4pt, right] at (5.2, 2.85) {\tiny c};
\draw (5, 3.1) rectangle (5.8, 3.3);
\node[above=4pt, right] at (5.2, 3.05) {\tiny e};
\draw (6.4, 2.5) rectangle (7.2, 2.7);
\node[above=4pt, right] at (6.6, 2.45) {\tiny b};
\draw (6.4, 2.7) rectangle (7.2, 2.9);
\node[above=4pt, right] at (6.6, 2.65) {\tiny c} ;	
\draw (6.4, 2.9) rectangle (7.2, 3.1);
\node[above=4pt, right] at (6.6, 2.85) {\tiny e};
\draw (6.4, 3.1) rectangle (7.2, 3.3);
\node[above=4pt, right] at (6.6, 3.05) {\tiny d};
\draw (7.4, 2.5) rectangle (8.2, 2.7);
\node[above=4pt, right] at (7.6, 2.45) {\tiny b};
\draw (7.4, 2.7) rectangle (8.2, 2.9);
\node[above=4pt, right] at (7.6, 2.65) {\tiny c} ;	
\draw (7.4, 2.9) rectangle (8.2, 3.1);
\node[above=4pt, right] at (7.6, 2.85) {\tiny d};
\draw (7.4, 3.1) rectangle (8.2, 3.3);
\node[above=4pt, right] at (7.6, 3.05) {\tiny e};
\draw[-, densely dotted]  (4.4, 2.5) -- (7.33, 1.9);
\draw[-stealth,black, densely dotted]  (5.4, 2.5) -- (7.33, 1.9);
\draw[-stealth,black, densely dotted]  (6.8, 2.5) -- (7.35, 1.9);
\draw[-stealth,black, densely dotted]  (7.8, 2.5) -- (7.34, 1.9);
\draw[-stealth,black]  (4.4, 2.5) -- (4.9, 1.9);
\draw[-stealth,black]  (5.4, 2.5) -- (4.9, 1.9);
\draw (4.5, 1.1) rectangle (5.3, 1.3);
\node[above=4pt, right] at (4.7, 1.05) {\tiny a};
\draw (4.5, 1.3) rectangle (5.3, 1.5);
\node[above=4pt, right] at (4.7, 1.25) {\tiny b} ;	
\draw (4.5, 1.5) rectangle (5.3, 1.7);
\node[above=4pt, right] at (4.7, 1.45) {\tiny c};
\draw (4.5, 1.7) rectangle (5.3, 1.9);
\node[above=4pt, right] at (4.7, 1.65) {};
\node[below] at (4.7, 1) {\tiny{(non-speculative or}};
\node[below] at (4.7, 0.75) {\tiny{aggressive merge strategy)}};
\draw (6.9, 1.1) rectangle (7.7, 1.3);
\node[above=4pt, right] at (7.1, 1.05) {\tiny b};
\draw (6.9, 1.3) rectangle (7.7, 1.5);
\node[above=4pt, right] at (7.1, 1.25) {\tiny c} ;	
\draw (6.9, 1.5) rectangle (7.7, 1.7);
\node[above=4pt, right] at (7.1, 1.45) {};
\draw (6.9, 1.7) rectangle (7.7, 1.9);
\node[above=4pt, right] at (7.1, 1.65) {};
\node[below] at (7.2, 1) {\tiny{(optimal merge strategy)}};
\end{tikzpicture}
}

%% file: fig/realex.tex
\scalebox{1.25}{\begin{tikzpicture}[fill=blue!20,font=\footnotesize]
\draw (0, -0.3) rectangle (1.8, 0.1);
\draw[fill=lightgray] (0, 0.1) rectangle (1.8, 0.3);
\node[above=-3pt] at (0.9, -0.3) {\tiny ref wd};
\node[above=-3pt] at (0.9, -0.1) {\tiny ref el};
\node[above=-3pt] at (0.9,  0.1) {\tiny bb1};
\draw (0, -1.2) rectangle (1.8, -1);
\draw[fill=lightgray] (0, -1) rectangle (1.8, -0.8);
\node[above=-3pt] at (0.9, -1) {\tiny bb2};
\node[above=-3pt] at (0.9, -1.2) {\tiny ref mil};
\draw (0, -2.7) rectangle (1.8, -1.7);
\draw[fill=lightgray] (0, -1.7) rectangle (1.8, -1.5);
\node[above=-3pt] at (0.9, -1.7) {\tiny bb3};
\node[above=-3pt] at (0.9, -1.93) {\tiny ref decis\_levl[mil]};
\node[above=-3pt] at (0.9, -2.1) {\tiny ref detl};
\node[above=-3pt] at (0.9, -2.3) {\tiny ref decis};
\node[above=-3pt] at (0.9, -2.5) {\tiny ref wd};
\node[above=-3pt] at (0.9, -2.7) {\tiny ref decis};
\draw (0, -4.4) rectangle (1.8, -4.2);
\draw[fill=lightgray] (0, -4.2) rectangle (1.8, -4);
\node[above=-3pt] at (0.9, -4.2) {\tiny bb5};
\node[above=-3pt] at (0.9, -4.4) {\tiny ref el};
\draw (-0.5, -3.7) rectangle (0.5, -3.3);
\draw[fill=lightgray]  (-0.5, -3.3) rectangle (0.5, -3.1);
\node[above=-3pt] at (0, -3.3) {\tiny bb4};
\node[above=-3pt] at (0, -3.5) {\tiny ref mil};
\node[above=-3pt] at (0, -3.7) {\tiny ref mil};
\draw (-1.5, -5.4) rectangle (0.6, -5);
\draw[fill=lightgray] (-1.5, -5) rectangle (0.6, -4.8);
\node[above=-3pt] at (-0.45, -5) {\tiny bb6};
\node[above=-3pt] at (-0.45, -5.24) {\tiny ref  quant26bt\_pos[mil]};
\node[above=-3pt] at (-0.45, -5.4) {\tiny ref ril};
\draw (1.2, -5.4) rectangle (3.3, -5);
\draw[fill=lightgray] (1.2, -5) rectangle (3.3, -4.8);
\node[above=-3pt] at (2.25, -5) {\tiny bb7};
\node[above=-3pt] at (2.25, -5.24) {\tiny ref  quant26bt\_neg[mil]};
\node[above=-3pt] at (2.25, -5.4) {\tiny ref ril};
\draw (0, -6.2) rectangle (1.8, -6);
\draw[fill=lightgray] (0, -6) rectangle (1.8, -5.8);
\node[above=-3pt] at (0.9, -6) {\tiny bb8};
\node[above=-3pt] at (0.9, -6.2) {\tiny ref ril};
\draw[-stealth,black] (0.9, -0.3) -- (0.9, -0.8);
\draw[-stealth,black] (0.9, -1.2) -- (0.9, -1.5);
\draw[-stealth,black] (0.9, -2.7) -- (0, -3.1);
\draw [-stealth,black] (0.9, -2.7)  to[out=-60,in=60]  (1, -4);
\draw [-stealth,black] (0, -3.7) -- (0.8, -4);
\draw[-stealth,black] (0.9, -4.4) -- (-0.45, -4.8);
\draw[-stealth,black] (0.9, -4.4) -- (2.25, -4.8);
\draw[-stealth,black] (-0.45, -5.4) -- (0.8, -5.8);
\draw[-stealth,black] (2.25, -5.4) -- (1, -5.8);
\draw [-stealth,black] (0, -3.7)  to[out=-135,in=-80] (-0.6,-3.7) to[out=100,in=-110] (-0.1,-0.8) to[out=70,in=135]  (0.8, -0.8);
\draw [red, dashed, -stealth] (0.3, -5.4)  to[out=-60,in=120] (1.5, -4.8);
\draw [red, -stealth] (1.5, -5.4)  to[out=-120,in=60] (0.3, -4.8);
\draw [red, dashed, -stealth] (1.5, -5.4)  to[out=-120,in=90] (0.9, -5.8);
\draw [red, -stealth] (0.3, -5.4)  to[out=-60,in=90] (0.9, -5.8);
\end{tikzpicture}}

%% file: fig/widening.tex
\scalebox{1.2}{
\begin{tikzpicture}[fill=blue!20,font=\footnotesize]

  \draw[->] (2.5, 0.5) -- (2.5, 0.1);
  \draw (2.5, 0.1) -- (3, -0.1);
  \draw (2.5, 0.1) -- (2, -0.1);
  \draw (3, -0.1) -- (3, -1.7);

  \draw[->] (2, -0.1) -- (2, -0.4);
  \draw(2, -0.4) -- (1.5, -0.6);
  \draw (2, -0.4) -- (2.5, -0.6);
  \draw (2.5, -0.6) -- (2.5, -1.2);
  \draw (1.5, -0.6) -- (1.5, -1.2);
  \draw(2.5, -1.2) -- (2, -1.4);
  \draw (1.5, -1.2) -- (2, -1.4);
  \draw[->] (2, -1.4) -- (2, -1.7);

  \draw (2, -1.7) -- (1.2, -1.7);
  \draw[->] (1.2, -1.7) -- (1.2, -0.9);
  \draw (1.2, -0.9) -- (1.2, -0.1);
  \draw (1.2, -0.1) -- (2, -0.1);

  \draw (2, -1.7) -- (2.5, -1.9);
  \draw (2, -1.7) -- (2.5, -1.9);

  \draw (3, -1.7) -- (2.5, -1.9);
  \draw[->] (2.5, -1.9) -- (2.5, -2.2);

\node[circle,fill=black,inner sep=0pt,minimum size=3pt] at (2.5, 0.3) {};
  \node[circle,fill=black,inner sep=0pt,minimum size=3pt] at (2.5, -0.9) {};
  \node[circle,fill=black,inner sep=0pt,minimum size=3pt] at (1.5, -0.9) {};

  \node[right] at (2.5, 0.3) {\tiny a};
  \node[right] at (1.5, -0.9) {\tiny b};
  \node[left] at (2.5, -0.9) {\tiny c};
\end{tikzpicture}
}

%% file: fig/widening1_v2.tex
\scalebox{1.2}{
\begin{tikzpicture}[fill=blue!20,font=\footnotesize]
  \draw[->] (2, 1.3) -- (2, 0.9);
  \draw (2, 0.9) -- (1.5, 0.7);
  \draw (2, 0.9) -- (2.5, 0.7);
  \draw (2.5, 0.7) -- (2.5, 0.1);
  \draw (1.5, 0.7) -- (1.5, 0.1);
  \draw (2.5, 0.1) -- (2, -0.1);
  \draw (1.5, 0.1) -- (2, -0.1);
  \draw[->] (2, -0.1) -- (2, -0.4);
  \draw(2, -0.4) -- (1.5, -0.6);
  \draw (2, -0.4) -- (2.5, -0.6);
  \draw (2.5, -0.6) -- (2.5, -1.2);
  \draw (1.5, -0.6) -- (1.5, -1.2);
  \draw(2.5, -1.2) -- (2, -1.4);
  \draw (1.5, -1.2) -- (2, -1.4);
  \draw[->] (2, -1.4) -- (2, -1.7);
  \draw (2, -1.7) -- (1.5, -1.9);
  \draw (2, -1.7) -- (2.5, -1.9);
  \draw (2.5, -1.9) -- (2.5, -2.5);
  \draw (1.5, -1.9) -- (1.5, -2.5);
  \draw (2.5, -2.5) -- (2, -2.7);
  \draw (1.5, -2.5) -- (2, -2.7);
  \draw[->] (2, -2.7) -- (2, -3);
  \node at (2, -3.2)  {\Large ...};

  \node[circle,fill=black,inner sep=0pt,minimum size=3pt] at (2, 1.1) {};
  \node[circle,fill=black,inner sep=0pt,minimum size=3pt] at (1.5, 0.4) {};
  \node[circle,fill=black,inner sep=0pt,minimum size=3pt] at (2.5, 0.4) {};
  \node[circle,fill=black,inner sep=0pt,minimum size=3pt] at (1.5, -0.9) {};
  \node[circle,fill=black,inner sep=0pt,minimum size=3pt] at (2.5, -0.9) {};
  \node[circle,fill=black,inner sep=0pt,minimum size=3pt] at (1.5, -2.2) {};
  \node[circle,fill=black,inner sep=0pt,minimum size=3pt] at (2.5, -2.2) {};
  \node[right] at (2, 1.1) {\tiny a};
  \node[right] at (1.5, 0.4) {\tiny b};
  \node[left] at (2.5, 0.4) {\tiny c};
  \node[right] at (1.5, -0.9) {\tiny b};
  \node[left] at (2.5, -0.9) {\tiny c};
  \node[right] at (1.5, -2.2) {\tiny b};
  \node[left] at (2.5, -2.2) {\tiny c};
  \node[right] at (2.3, 1.1)   {\tiny [a, $\bot$, $\bot$, $\bot$]};
  \node[left]  at (1.4, 0.4)   {\tiny [b, a, $\bot$, $\bot$]};
  \node[right] at (2.6, 0.4)   {\tiny [c, a, $\bot$, $\bot$]};
  \node[left]  at (1.4,  -0.9) {\tiny [b, $\{\}$, a, $\bot$]};
  \node[right] at (2.6,  -0.9) {\tiny [c, $\{\}$, a, $\bot$]};
  \node[left]  at (1.4, -2.2)  {\tiny [b, $\{\}$, $\{\}$, a]};
  \node[right] at (2.6, -2.2)  {\tiny [c, $\{\}$, $\{\}$, a]};

  \node[right] at (2.3, -0.25) {\tiny [$\{\}$, a, $\bot$, $\bot$]};
  \node[right] at (2.3, -1.55) {\tiny [$\{\}$, $\{\}$, a, $\bot$]};
  \node[right] at (2.3, -2.85) {\tiny [$\{\}$, $\{\}$, $\{\}$, a]};

  \draw[dotted] (0.2, 0.7) -- (3.8, 0.7);
  \draw[dotted] (0.2, 0.1) -- (3.8, 0.1);
  \draw[dotted] (0.2, -0.6) -- (3.8, -0.6);
  \draw[dotted] (0.2, -1.2) -- (3.8, -1.2);
  \draw[dotted] (0.2, -1.9) -- (3.8, -1.9);
  \draw[dotted] (0.2, -2.5) -- (3.8, -2.5);

\end{tikzpicture}
}

%% file: fig/newcachejoin_v2.tex
  \begin{tikzpicture}[fill=blue!20,font=\footnotesize]
  \draw (0, 0) rectangle (1.2, 0.3);
  \node[above=4pt, right] at (0.4, 0)   {k};
  \draw (0, 0.3) rectangle (1.2, 0.6);
  \node[above=4pt, right] at (0.4, 0.3) {z} ;	
  \draw (0, 0.6) rectangle (1.2, 0.9);
  \node[above=4pt, right] at (0.4, 0.6) {y};
  \draw (0, 0.9) rectangle (1.2, 1.2);
  \node[above=4pt, right] at (0.4, 0.9) {x};
  \draw (2, 0) rectangle (3.2, 0.3);
  \node[above=4pt, right] at (2.4, 0)   {k};
  \draw (2, 0.3) rectangle (3.2, 0.6);
  \node[above=4pt, right] at (2.4, 0.3) {x} ;	
  \draw (2, 0.6) rectangle (3.2, 0.9);
  \node[above=4pt, right] at (2.4, 0.6) {z};
  \draw (2, 0.9) rectangle (3.2, 1.2);
  \node[above=4pt, right] at (2.4, 0.9) {t};
  \draw (1.0, -1) rectangle (2.2, -0.7);
  \node[above=4pt, right] at (1.05, -1)    {\{\textcolor{purple}{$\exists$x, $\exists$t}\}};
  \draw (1, -1.3) rectangle (2.2, -1);
  \node[above=4pt, right] at (1.05, -1.3)  {\{\textcolor{purple}{$\exists$y, $\exists$z}\}} ;	
  \draw (1, -1.6) rectangle (2.2, -1.3);
  \node[above=4pt, right] at (1.2, -1.6)  {\{x, z\}};
  \draw (1, -1.9) rectangle (2.2, -1.6);
  \node[above=4pt, right] at (1.1, -1.9)  {\{\textcolor{purple}{$\exists$k,} k\}};
  \draw (3.5, -1) rectangle (4.7, -0.7);
  \node[above=4pt, right] at (3.6, -1)    {\{\textcolor{purple}{$\exists$y,} y\}};
  \draw (3.5, -1.3) rectangle (4.7, -1);
  \node[above=4pt, right] at (3.55, -1.3)  {\{\textcolor{purple}{$\exists$x, $\exists$t}\}} ; 
  \draw (3.5, -1.6) rectangle (4.7, -1.3);
  \node[above=4pt, right] at (3.7, -1.6)  {\{\textcolor{purple}{$\exists$z}\}};
  \draw (3.5, -1.9) rectangle (4.7, -1.6);
  \node[above=4pt, right] at (3.5, -1.9)  {\{\textcolor{purple}{$\exists$k,} x, z\}};
  \draw[-stealth,black] (0.6, 0) -- (1.4, -0.7);
  \draw[-stealth,black] (2.6, 0) -- (1.8, -0.7);
  \draw[-stealth,black] (3.4, 0.15) -- (3.4, 1.05) node [align =center,pos = 0.5,right] {young\\age};
  \draw[-stealth,black] (2.3, -1.3) -- (3.4, -1.3);
  \node[above=4pt, right] at (2.5, -1.3) {ref y} ; 
  \end{tikzpicture}
 

%% file: fig/exist_hit_v2.tex
\scalebox{1}{
\begin{tikzpicture}[fill=blue!20,font=\footnotesize]
  \draw[->] (2, 1.3) -- (2, 0.9);
  %\draw (2, 1.3) -- (2, 0.9);
  \draw (2, 0.9) -- (1.5, 0.7);
  \draw (2, 0.9) -- (2.5, 0.7);
  \draw (2.5, 0.7) -- (2.5, 0.1);
  \draw (1.5, 0.7) -- (1.5, 0.1);
  \draw (2.5, 0.1) -- (2, -0.1);
  \draw (1.5, 0.1) -- (2, -0.1);
  \draw[->] (2, -0.1) -- (2, -0.4);
  \draw(2, -0.4) -- (1.5, -0.6);
  \draw (2, -0.4) -- (2.5, -0.6);
  \draw (2.5, -0.6) -- (2.5, -1.2);
  \draw (1.5, -0.6) -- (1.5, -1.2);
  \draw(2.5, -1.2) -- (2, -1.4);
  \draw (1.5, -1.2) -- (2, -1.4);
  \draw[->] (2, -1.4) -- (2, -1.7);
  \draw (2, -1.7) -- (1.5, -1.9);
  \draw (2, -1.7) -- (2.5, -1.9);
  \draw (2.5, -1.9) -- (2.5, -2.5);
  \draw (1.5, -1.9) -- (1.5, -2.5);
  \draw (2.5, -2.5) -- (2, -2.7);
  \draw (1.5, -2.5) -- (2, -2.7);
  \draw[->] (2, -2.7) -- (2, -3);
  \node[circle,fill=black,inner sep=0pt,minimum size=3pt] at (2, 1.1) {};
  \node[circle,fill=black,inner sep=0pt,minimum size=3pt] at (1.5, 0.4) {};
  \node[circle,fill=black,inner sep=0pt,minimum size=3pt] at (2.5, 0.4) {};
    \node[circle,fill=black,inner sep=0pt,minimum size=3pt] at (1.5, -0.9) {};
  \node[circle,fill=black,inner sep=0pt,minimum size=3pt] at (2.5, -0.9) {};
    \node[circle,fill=black,inner sep=0pt,minimum size=3pt] at (1.5, -2.2) {};
  \node[circle,fill=black,inner sep=0pt,minimum size=3pt] at (2.5, -2.2) {};
  \node[right] at (2, 1.1) {\tiny a};
  \node[left] at (1.5, 0.4) {\tiny b};
  \node[right] at (2.5, 0.4) {\tiny c};
    \node[left] at (1.5, -0.9) {\tiny b};
  \node[right] at (2.5, -0.9) {\tiny c};
    \node[left] at (1.5, -2.2) {\tiny b};
  \node[right] at (2.5, -2.2) {\tiny c};
  \node[right] at (2.3, 1.1)   {\tiny [\{\textcolor{purple}{$\exists$a,} a\}, $\bot$, $\bot$, $\bot$]};
  \node[left] at (1.2, 0.4)    {\tiny [\{\textcolor{purple}{$\exists$b,} b\}, \{\textcolor{purple}{$\exists$a,} a\}, $\bot$, $\bot$]};
  \node[right] at (2.8, 0.4)   {\tiny [\{\textcolor{purple}{$\exists$c,} c\}, \{\textcolor{purple}{$\exists$a,} a\}, $\bot$, $\bot$]};
    \node[left] at (1.2,  -0.9){\tiny [\{\textcolor{purple}{$\exists$b,} b\}, \{\textcolor{purple}{$\exists$a, $\exists$c}\}, a, $\bot$]};
  \node[right] at (2.8,  -0.9) {\tiny [\{\textcolor{purple}{$\exists$c,} c\}, \{\textcolor{purple}{$\exists$a, $\exists$b}\}, a, $\bot$]};
    \node[left] at (1.2, -2.2) {\tiny [\{\textcolor{purple}{$\exists$b,} b\}, \{\textcolor{purple}{$\exists$a, $\exists$c}\}, a, $\bot$]};
  \node[right] at (2.8, -2.2)  {\tiny [\{\textcolor{purple}{$\exists$c,} c\}, \{\textcolor{purple}{$\exists$a, $\exists$b}\}, a, $\bot$]};

  \node[right] at (2.3, -0.25) {\tiny [\{\textcolor{purple}{$\exists$b, $\exists$c}\}, \{\textcolor{purple}{$\exists$a,} a\}, $\bot$, $\bot$]};
  \node[right] at (2.3, -1.55) {\tiny [\{\textcolor{purple}{$\exists$b, $\exists$c}\}, \{\textcolor{purple}{$\exists$a}\}, a, $\bot$]};
  \node[right] at (2.3, -2.85) {\tiny [\{\textcolor{purple}{$\exists$b, $\exists$c}\}, \{\textcolor{purple}{$\exists$a}\}, a, $\bot$]};

   \draw[dotted] (0, 0.7) -- (4, 0.7);
   \draw[dotted] (0, 0.1) -- (4, 0.1);
   \draw[dotted] (0, -0.6) -- (4, -0.6);
   \draw[dotted] (0, -1.2) -- (4, -1.2);
   \draw[dotted] (0, -1.9) -- (4, -1.9);
    \draw[dotted] (0, -2.5) -- (4, -2.5);
\end{tikzpicture}
}